\documentclass{elsart}
\usepackage{graphics}
\usepackage{psfig}
\usepackage{color}

\def\deg{{\rm deg}}

\def\qed{\hfill  \framebox(5,5){}}
\def\lcoeff{{\rm lcoeff}}
\def\coeff{{\rm coeff}}

\def\Res{{\rm Res}}

\expandafter\chardef\csname pre amssym.def
at\endcsname=\the\catcode`\@ \catcode`\@=11

\def\undefine#1{\let#1\undefined}
\def\newsymbol#1#2#3#4#5{\let\next@\relax
 \ifnum#2=\@ne\let\next@\msafam@\else
 \ifnum#2=\tw@\let\next@\msbfam@\fi\fi
 \mathchardef#1="#3\next@#4#5}
\def\mathhexbox@#1#2#3{\relax
 \ifmmode\mathpalette{}{\m@th\mathchar"#1#2#3}%
 \else\leavevmode\hbox{$\m@th\mathchar"#1#2#3$}\fi}
\def\hexnumber@#1{\ifcase#1 0\or 1\or 2\or 3\or 4\or 5\or 6\or 7\or 8\or
 9\or A\or B\or C\or D\or E\or F\fi}

\font\tenmsa=msam10 \font\sevenmsa=msam7 \font\fivemsa=msam5
\newfam\msafam
\textfont\msafam=\tenmsa \scriptfont\msafam=\sevenmsa
\scriptscriptfont\msafam=\fivemsa
\edef\msafam@{\hexnumber@\msafam} \mathchardef\dabar@"0\msafam@39
\def\dashrightarrow{\mathrel{\dabar@\dabar@\mathchar"0\msafam@4B}}
\def\dashleftarrow{\mathrel{\mathchar"0\msafam@4C\dabar@\dabar@}}

        \font\tenmsb=msbm10
\font\sevenmsb=msbm7 \font\fivemsb=msbm5
\newfam\msbfam
\textfont\msbfam=\tenmsb \scriptfont\msbfam=\sevenmsb
\scriptscriptfont\msbfam=\fivemsb
\edef\msbfam@{\hexnumber@\msbfam}
\def\Bbb#1{\fam\msbfam\relax#1}

\newtheorem{theorem}{{\bf Theorem}}
\newtheorem{remark}{{\bf Remark}}
\newtheorem{definition}[theorem]{{\bf Definition}}
\newtheorem{corollary}[theorem]{{\bf Corollary}}
\newtheorem{proposition}[theorem]{{\bf Proposition}}
\newtheorem{lemma}[theorem]{{\bf Lemma}}
\newtheorem{example}{{\bf Example}}

\begin{document}
\begin{frontmatter}



\title{On the Different Shapes Arising in a Family of Rational Curves Depending on a Parameter}

\author[a]{Juan Gerardo Alcazar\thanksref{proy}},
\ead{juange.alcazar@uah.es}

\address[a]{Departamento de Matem\'aticas, Universidad de Alcal\'a,
E-28871-Madrid, Spain}

\thanks[proy]{Author supported by the Spanish Supported `` Ministerio de
Ciencia e Innovacion" under the Project MTM2008-04699-C03-01.}

\begin{abstract}
Given a family of rational curves depending on a real parameter, defined by its parametric equations, we provide an algorithm
to compute a finite partition of the parameter space (${\Bbb R}$, in general) so that the shape of the
family stays invariant along each element of the partition. So, from this partition the topology types in the
family can be determined. The algorithm is based on a geometric interpretation of
previous work (\cite{JGRS}) for the implicit case. However, in our case the algorithm works directly with the
parametrization of the family, and the implicit equation does not
need to be computed. Timings comparing the algorithm in the implicit and the parametric cases are
given; these timings show that the parametric algorithm developed here provides in general better results than the known algorithm for the
implicit case.
\end{abstract}
\end{frontmatter}

\section{Introduction}\label{section-introduction}

Given a family ${\mathcal F}$ of algebraic curves depending on a
real parameter, it is clear that the shape of the curves in the
family may change as the value of the parameter changes. In the
C.A.G.D. context, one has a good example of this phenomenon in the
family of {\it offset} curves to a given curve (see \cite{ASS96},
\cite{L95a}, \cite{Po95}, \cite{PP98b}), where the parameter is
the offsetting distance. Take for example the well-known case of
the offset to the parabola (see \cite{Far1}). In this case three
different shapes, which can be seen in Figure 1 (together with the
original parabola, in thinner line), arise; here, one may observe
that for distances $d<1/2$, the offsets show no cusps, but they
exhibit an isolated point (i.e. a {\it geometric extraneous
component}), and that for $d>1/2$, two cusps and a
self-intersection arise.

\begin{figure}[ht]
\begin{center}
\centerline{$\begin{array}{ccc}
\psfig{figure=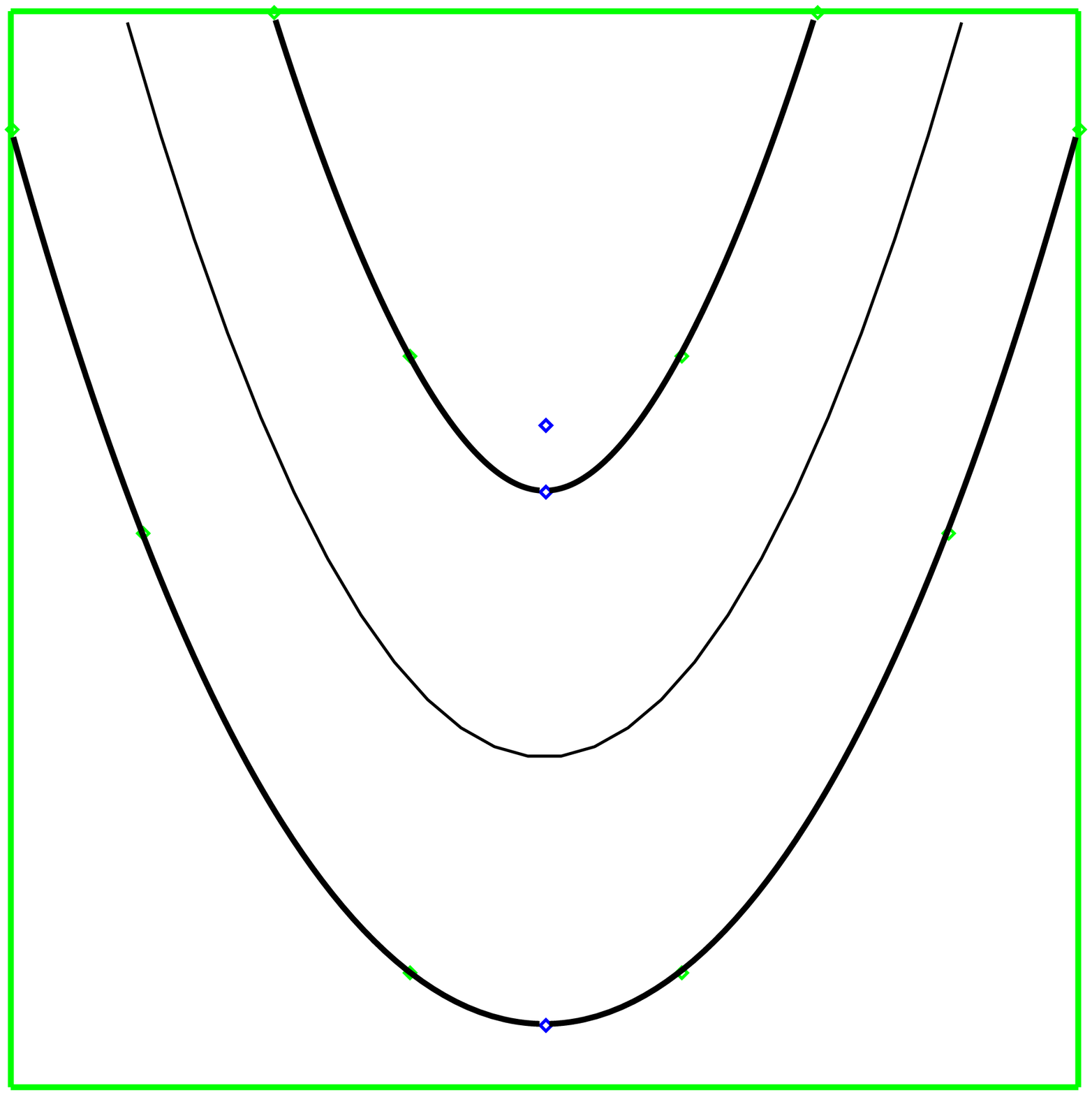,width=4cm,height=3cm} &
\psfig{figure=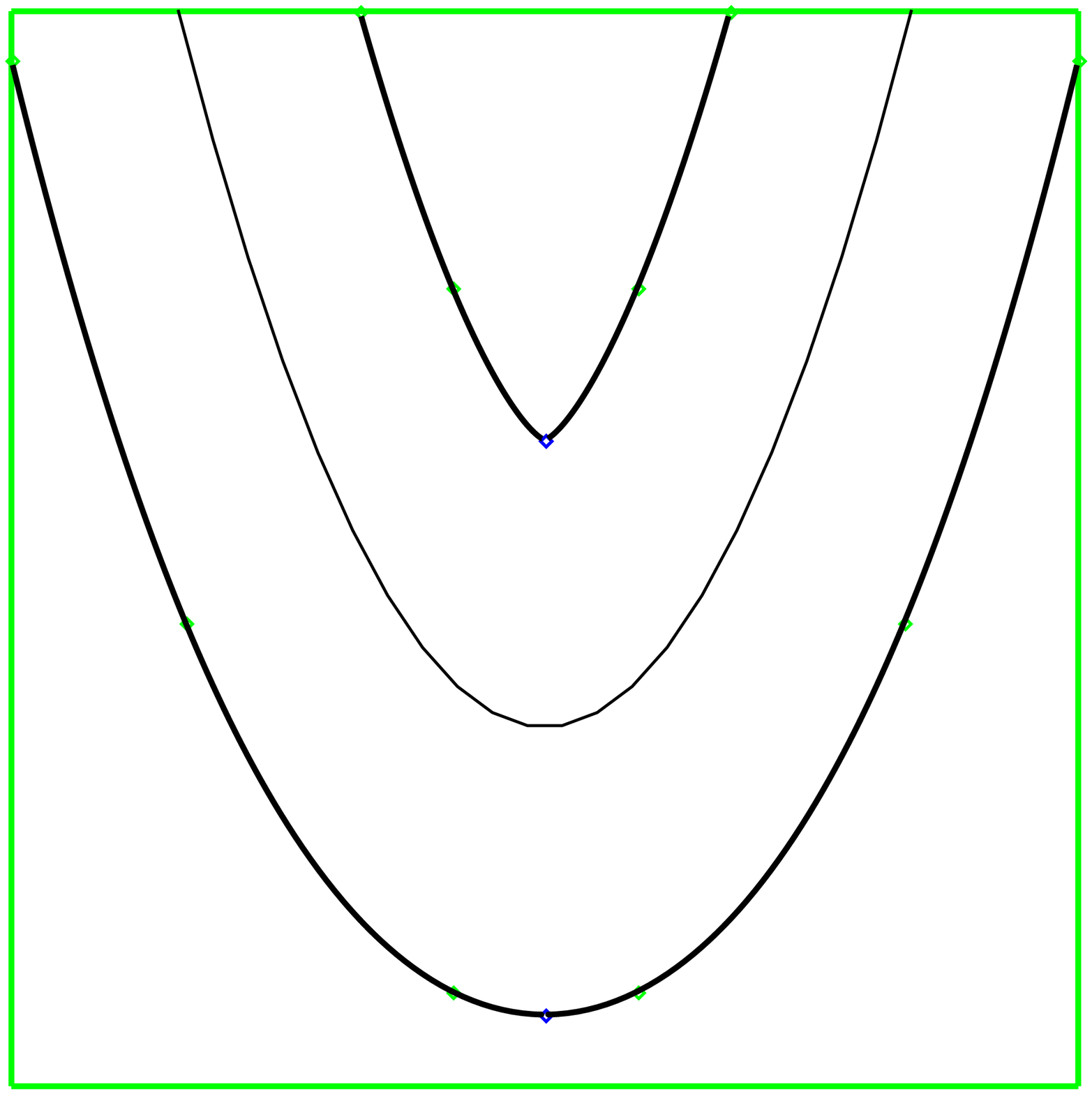,width=4cm,height=3cm} &
\psfig{figure=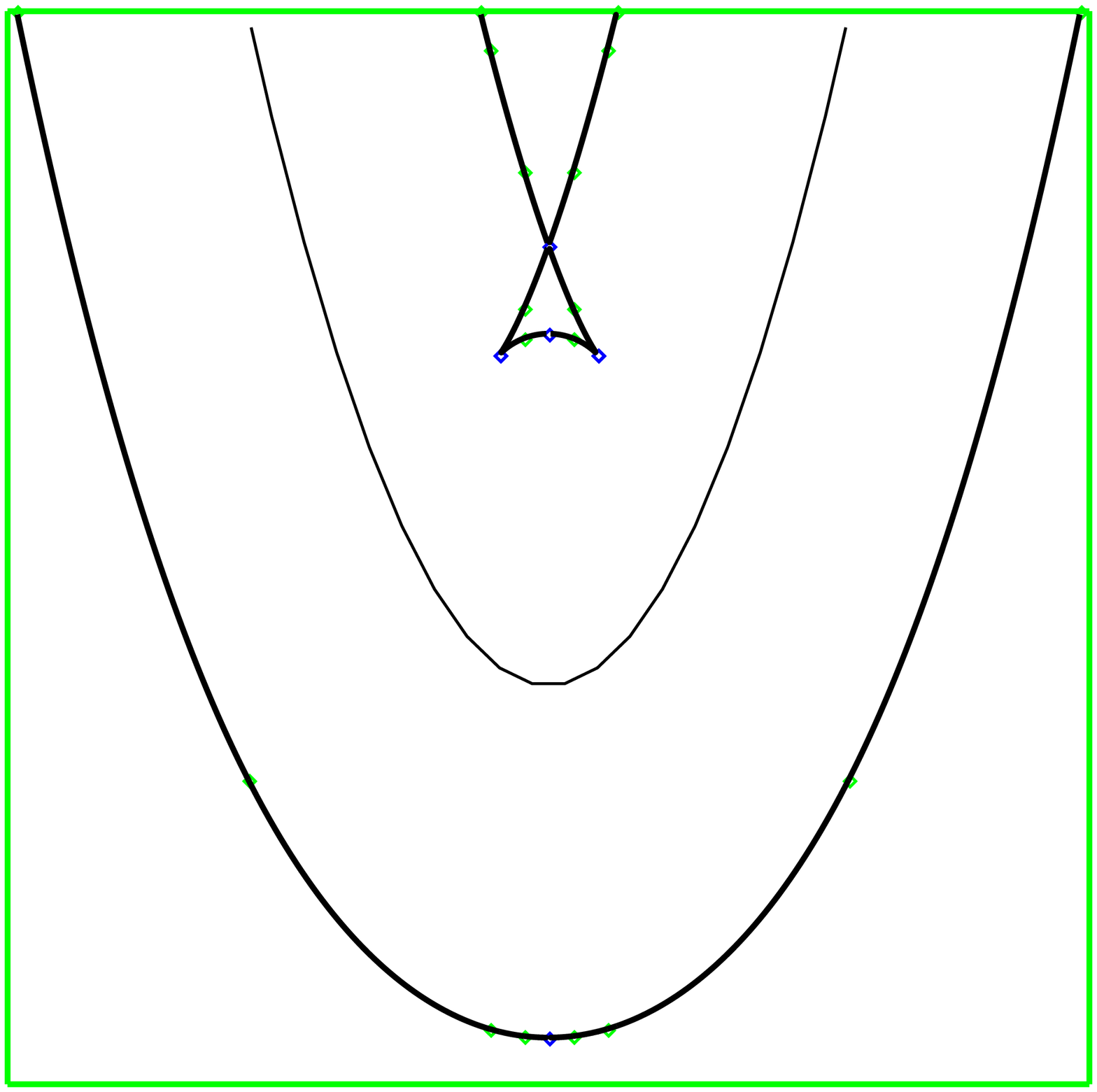,width=4cm,height=3cm}
\end{array}$}
\end{center}

\caption{Topology types of the offsets to $y^2-x=0$: $d<1/2$
(left); $d=1/2$ (center); $d>1/2$ (right)}. In the middle of each
figure, the parabola has been also plotted.
\end{figure}

The problem of computing the different shapes arising in a family
of algebraic curves depending on a parameter has been considered
in \cite{JGRS}, \cite{Mourrain-2}, under different perspectives
but reaching similar results, for the case when the family is
implicitly given; so, in these papers the family to be analyzed is
defined as the zero set of a polynomial $F(x,y,\lambda)$, where
$\lambda$ is regarded as a real parameter. In particular, using
the result in \cite{JGRS}, the topology types in Figure 1 can be
determined. For this purpose, the strategy is the following.
First, one computes a finite set ${\mathcal A}$ so that in between
two consecutive elements of ${\mathcal A}$, the topology type of
the family does not change; we refer to such a set as a {\it
critical set}; in the case of the offset to the parabola,
${\mathcal A}=\{1/2\}$. Then, this critical set induces a finite
partition of the parameter space; in the above example, the
partition is $(0,1/2)$, $\{1/2\}$, $(1/2,\infty)$ (notice that the
offsetting distance is a positive real number). Finally, taking a
representative for each element of the partition, and applying if
necessary standard methods for describing the topology of a plane
algebraic curve (see \cite{Eigen}, \cite{Lalo}, \cite{Hong}), the
different shapes in the family can be computed.

In this paper, starting from the results in \cite{JGRS} we address the same problem but for the case,
specially interesting in the C.A.G.D. context, of a family ${\mathcal F}$ of
rational curves depending on a parameter, defined in parametric form. Thus, our input is
\[
\left \{ \begin{array}{l}
x=u(t,\lambda)\\
y=v(t,\lambda)\\
\end{array} \right.
\]
where $u(t,\lambda), v(t,\lambda)$ are rational functions in terms of $t,\lambda$, and $\lambda\in {\Bbb R}$ is a parameter.
In these conditions, we consider the computation of a critical set of ${\mathcal F}$, but without computing or making use of the
implicit equation of the family. So, the main result of the paper is an algorithm for carrying out this computation.

The algorithm is advantageous when a rational parametrization of the family to be analyzed is available. In fact, in our
experimentation we have found several examples (see the comparison table in Subsection \ref{exp-res}) of families
 that are hard to study with the result in \cite{JGRS} (i.e.
the computations are too heavy in implicit form), but which can be analyzed with the algorithm provided here.

The algorithm is based on a geometrical interpretation of the
results in \cite{JGRS}. This interpretation suggests a certain
geometrical process to be performed in order to compute a critical
set. Hence, the question is to carry out that process in
parametric form. The geometric analysis of the main results in
\cite{JGRS} can be found in Section 2. In Section 3, we show how
to check different hypotheses that we impose on the
parametrization, and how to ``prepare" the family before applying
the algorithm; the ideas in this section are illustrated in
Example \ref{prep-off-cardiod}, where the offset family to a
cardioid is considered. In Section 4, we provide the full
algorithm, we complete the analysis of Example
\ref{prep-off-cardiod}, and we give a comparison table showing the
timings of our algorithm, compared to those corresponding to the
algorithm in \cite{JGRS}; these timings show that our algorithm
yields in general better results than the known algorithm for the
implicit case. Finally, in Section \ref{sec-conclusions} we
present the conclusions of our study. The parametrizations used
for comparing timings are given in Appendix I.

\section{Preliminaries.}\label{section-geom-interpret}

\subsection{Known Results for the Implicit Case.}\label{crucial-section}

Let $F\in {\Bbb R}[x,y,\lambda]$. For all $\lambda_0\in {\Bbb R}$
such that $F(x,y,\lambda_0)$ is not identically $0$ we have that
$F(x,y,\lambda_0)=F_{\lambda_0}(x,y)$ defines an algebraic curve.
So, we can say that $F$ defines a family ${\mathcal F}$ of
algebraic curves algebraically depending on the parameter
$\lambda$. By Hardt's Semialgebraic Triviality Theorem (see
Theorem 5.46 in \cite{Basu}), it holds that the number of topology
types of ${\mathcal F}$ (i.e. the different shapes arising in the
family as $\lambda_0$ moves in ${\Bbb R}$) must be finite.
Therefore, it makes sense to consider the problem of computing the
topology types arising in the family. This problem has been
addressed in \cite{JGRS}, for the case when the family is
implicitly defined. So, in this subsection we recall the main
aspects (hypotheses, notation, and results) of this paper.

More precisely, given $F(x,y,\lambda)$ one can associate an algebraic surface
$S$ with the family ${\mathcal F}$ defined by $F$ by substituting $\lambda:=z$ in $F$; thus, the members of ${\mathcal F}$
are the {\it level curves}
of $S$, i.e. the sections of $S$ with planes normal to the $z$-axis. So, the problem of computing the topology types in
${\mathcal F}$ can be re-interpreted as the computation of the
topology types arising in the family of level curves of $S$. This is exactly the question
addressed in \cite{JGRS}. In order to solve this
problem, in \cite{JGRS} the following definition is introduced.

\begin{definition} \label{def-crit-set}
Let $S$ be an algebraic surface. We say that ${\mathcal A}\subset {\Bbb R}$ is a {\sf critical set} of $S$, if it is finite and it contains all the $z$-values where the topology type of the level curves of $S$ changes; i.e., if the topology type of the level curves of $S$ stays invariant along any interval delineated by two consecutive elements of ${\mathcal A}$.
\end{definition}

We speak about the critical set of ${\mathcal F}$, to mean the critical set of the surface associated with the family.
Notice that if a critical set ${\mathcal A}=\{a_1,\ldots,a_r\}$ is computed, then the parameter space (${\Bbb R}$, in this case)
can be decomposed as \[(-\infty,a_1)\cup \{a_1\}\cup (a_1,a_2)\cup \cdots \cup \{a_r\}\cup (a_r,+\infty)\]Then, taking a representative
for each element of the above partition, and applying standard methods (\cite{Eigen}, \cite{Lalo}, \cite{Hong}) for describing the
topology of an algebraic curve, the topology types in the family can be computed. Hence, the crucial question is the computation of
a critical set. This is the problem addressed in \cite{JGRS}. In the rest of the subsection, we recall the hypotheses, notation (that
we will also follow here) and
main result of \cite{JGRS}.

\underline{Hypotheses:} The hypotheses imposed in \cite{JGRS} on the surface $S$ to be analyzed, are: (a) $F$ is square-free, and
depends on the variable $y$; (b) $F$ does not contain any factor only depending on the variable $z$ (i.e. $S$ has no component
normal to the $z$-axis); (c) the leading coefficient of $F$ w.r.t. the variable $y$, does not depend on $x$.

\underline{Notation:} Given a polynomial $G$, $\sqrt{G}$ denotes
the square-free part of $G$, i.e. the product of all the
irreducible factors of $G$ taken with multiplicity 1. Also,
$D_w(G)$ denotes the {\it discriminant} of $G$ with respect to the
variable $w$, i.e. the resultant of $G$ and its partial derivative
with respect to $w$. We write this resultant as
$D_w(G)=\Res_w(G,\frac{\partial G}{\partial w})$. Furthermore, the
following polynomials are introduced:

\[
\begin{array}{lr}
M(x,z):=
\sqrt{D_y(F)}&\mbox{,    } R(z):=\left\{\begin{array}{cll} 0 & \,\, & \mbox{if }\deg_x(M)=0
\\ D_x(M(x,z)) & \,\, & \mbox{otherwise}
\end{array} \right.
\end{array}
\]

\underline{Result:} the main result of \cite{JGRS} is the following (see Theorem 4 and Theorem 13 in \cite{JGRS}):

\begin{theorem} \label{main-theorem-top}
Let $S$ be an algebraic surface implicitly defined by $F\in {\Bbb R}[x,y,z]$, fulfilling the above hypotheses. Then the following
statements hold:
\begin{itemize}
\item [(1)] If $R$ is not identically zero, then the set
of real roots of $R$ is a critical set of $S$. If $R$ has no real
roots, then the elements of the family show just one topology
type.
\item [(2)] If $R$ is identically zero, then $M=M(z)$, and the set of real roots
of $M$ is a critical set of $S$.
\end{itemize}
\end{theorem}


\subsection{Geometrical interpretation of the results for the implicit case.}\label{important-obs}

Let us provide a geometrical interpretation of the process,
according to Theorem \ref{main-theorem-top}, giving rise to a
critical set of $S$. For this purpose, let ${\mathcal C}=V(F,F_y)$
be the algebraic variety defined by $F$ and its partial derivative
$F_y$ (i.e. the intersection of the surfaces defined by $F$ and
$F_y$, respectively). Under the assumptions made on $S$, one may
check that $\gcd(F,F_y)=1$; so, ${\mathcal C}$ has dimension 1,
i.e. it is a space algebraic curve. Taking into account that the
normal direction to the surface $S$ at each point $P\in S$ is
defined by the gradient vector $\nabla
F(P)=(F_x(P),F_y(P),F_z(P))$, one may see that ${\mathcal C}$
consists of the points of $S$ where the normal vector is either
parallel to the $xz$-plane, or identically zero (i.e.
singularities of $S$). Also, let us denote the curve defined by
the polynomial $M(x,z)$ on the $xz$-plane, as ${\mathcal M}$. Now,
from a geometric point of view, in order to compute a critical set
by means of Theorem \ref{main-theorem-top}, one has to perform two
different phases:

\begin{itemize}
\item [(1)] \underline{\it Computation of ${\mathcal M}$ (Projection Phase):} from
basic properties of resultants (see for example p. 255 in \cite{SWPD}), one may see that the curve
${\mathcal M}$ consists of the
projection onto the $xz$-plane of ${\mathcal C}$, together with
the curve defined by the leading coefficient w.r.t. $y$ of $F$, denoted as $\lcoeff_y(F)$;
since by hypothesis $\lcoeff_y(F)$ does not depend on $x$,
this last curve consists of finitely many lines $z-z_0=0$ where
$z_0$ is a root of $\lcoeff_y(F)$.\\

\item [(2)] \underline{\it Computation of the critical set (Analysis of the Projection):} the
following real $z$-values must be computed:

\begin{itemize}
\item[(i)] the $z$-coordinates of the points
of ${\mathcal M}$ with tangent parallel to the $x$-axis (included the values
$z_a$'s so that $z-z_a$ is a component of ${\mathcal M}$).
\item[(ii)]
the $z$-coordinates of the singularities of ${\mathcal M}$
\item[(iii)] the values $z_b$'s so that $z-z_b$ is a horizontal asymptote of ${\mathcal M}$
\end{itemize}

In the case $M=M(z)$ this is clear. In other case, one must notice that from well-known properties of resultants (again, p. 255 in \cite{SWPD}), the roots of
$\Res_x(M,M_x)$ correspond either to the $z$-coordinates of the solutions of the polynomial
system $\{M(x,z)=0,M_x(x,z)=0\}$, or to the roots of $\lcoeff_x(M)$. Then, it suffices to interpret the solutions of the system, and the real roots of $\lcoeff_x(M)$, from a geometric point of view.
    \end{itemize}

 In the rest of the paper, we will refer to these $z$-values as {\sf $(1)$-values}, {\sf $(2)$-values} and {\sf $(3)$-values}, respectively.

\subsection{Notation and Hypotheses for the parametric case} \label{subsec-statement}
In the rest of the paper, let ${\mathcal F}$ be a family of rational algebraic curves, algebraically depending on a parameter $\lambda$, defined by the parametric equations
\[
\left \{ \begin{array}{l}
x=u(t,\lambda)\\
y=v(t,\lambda)\\
\end{array} \right.
\]
where $u,v$ are real, rational functions of the variables $t,\lambda$ in reduced form (i.e. the numerator and denominator of $u,v$ share
no common factor), not identically $0$. Thus, for almost all real values of $\lambda$ the above equations define a rational curve of
parameter $t$. In fact, the only exceptions are the values of $\lambda$ causing that some  denominator of $u,v$ vanishes. Now our purpose
is to study the topology types arising in ${\mathcal F}$ as $\lambda$ moves in ${\Bbb R}$. In our approach, the key for computing these
topology types is the computation of
 a finite partition of ${\Bbb R}$ so that for each element of the partition, the topology type of the family stays invariant. Hence, in
 the sequel we focus on this question. For this purpose, observe that the equations
\[
\left \{ \begin{array}{l}
x=u(t,\lambda)\\
y=v(t,\lambda)\\
z=\lambda\\
\end{array} \right.
\]
define a rational surface $S$ in parametric form, whose level curves are exactly the members of the family ${\mathcal F}$. Hence, our aim is the computation of a critical set of $S$. For this purpose, we introduce the following notation:
\begin{itemize}
\item $\phi_{\lambda}(t)=(u(t,\lambda),v(t,\lambda))$ is the parametrization of the family ${\mathcal F}$. Hence, the associated surface $S$ is defined by the parametrization $(u(t,\lambda),v(t,\lambda),\lambda)$.
    \item $F_\lambda(x,y)=F(x,y,\lambda)$ defines the implicit equation of ${\mathcal F}$. Hence, the implicit equation of $S$ is $F(x,y,z)=0$.
    \end{itemize}

Now in order to make precise the hypotheses that we require on the
family ${\mathcal F}$, we need to recall the notion of {\sf proper
parametrization}. One says that a parametrization $\phi(t)$ of a
rational curve ${\mathcal H}$ is {\sf proper}, if there are just
finitely many points of ${\mathcal H}$ generated simultaneously by
several different values of the parameter $t$; intuitively
speaking, this means that the curve is traced just once as $t$
moves in ${\Bbb R}$. The interested reader may find more
information on this topic in Chapter 4.2 of \cite{SWPD}. Then, we
consider the following hypotheses:

\underline{Hypotheses:}
\begin{itemize}
\item [($H_1$)] The parametrization $\phi_{\lambda}(t)=(u(t,\lambda),v(t,\lambda))$ is proper for almost all values of $\lambda$ (i.e. there are just finitely many values of $\lambda$ such that $\phi_{\lambda}(t)$ is not proper).
    \item [($H_2$)] $\deg_y(F_{\lambda})=\deg(F_{\lambda})$, i.e. considering $F_{\lambda}$ as a polynomial in the variables $x,y$, the degree of $F_{\lambda}$ w.r.t. the variable $y$ equals the total degree of $F_{\lambda}$; in particular, the leading coefficient of $F_{\lambda}$ w.r.t. $y$ does not depend on $x$, though it may depend on $\lambda$.
        \item [($H_3$)] The function $u(t,\lambda)$ depends on the variable $t$.
            \end{itemize}

        In the following, we will refer to these hypotheses as ($H_1$) and ($H_2$), respectively.
        Observe that ($H_2$)
        is slightly more restrictive than the hypothesis (c) required in the implicit case.
        On the other hand, ($H_3$) guarantees that $F_{\lambda}$ (and therefore $F$) depends on the variable $y$.
        Notice that if ($H_3$) does not hold, the analysis is reduced to the family $x=u(\lambda)$ (consisting just of
        finitely many lines normal to the $x$-axis), and the problem is trivial. Finally, since $S$ is a rational surface then it is irreducible;
        in particular, $S$ cannot represent a
        plane normal to the $z$-axis because from the parametrization it is clear that the $z$-coordinate cannot be constant. Thus, whenever
        the above hypotheses hold, the surface $S$ described by the considered parametrization
        satisfies all the hypotheses required in the implicit version of the
        problem.

The problem of checking the above
hypotheses ($H_1$) and ($H_2$) is addressed in the next section. Checking ($H_3$) is trivial; so, in the sequel we assume that ($H_3$) holds.

\section{Checking Hypotheses}\label{sec-checking}

In this section we provide an algorithm for checking the
hypotheses ($H_1$) and ($H_2$) introduced in Subsection \ref{subsec-statement}. In addition, the
algorithm addressed here provides also a list of finitely many ``special" values of the parameter $\lambda$,
which will be important (see the next section) in order to compute a critical set of the surface $S$ associated with the family.

\subsection{Checking hypothesis ($H_1$)}\label{subsec-uno}

In this subsection we consider the problem of checking whether ($H_1$) holds, or not. For this purpose, we start with a technical lemma on the behavior of the $\gcd$ of two polynomials of ${\Bbb R}[x,y,\lambda]$, under the specialization of the parameter $\lambda$. Hence, let $\varphi_a$ be the natural homomorphism of ${\Bbb R}[x,y,\lambda]$ into ${\Bbb R}[x,y]$, i.e. for $a\in {\Bbb R}$,
\[
\begin{array}{rcl}
\varphi_a:{\Bbb R}[x,y,\lambda] & \to & {\Bbb R}[x,y]\\
f(x,y,\lambda) & \to & \varphi_a(f)=f(x,y,a)
\end{array}
\]
Moreover, if $f,g\in {\Bbb R}[x,y,\lambda]$, we write $f=\bar{f}\cdot \gcd(f,g)$, $g=\bar{g}\cdot \gcd(f,g)$, and we define
the following sets:
\[
\begin{array}{c}
{\mathcal B}_1^{(x,y)}(f,g)=\{a\in {\Bbb R}|\varphi_a\left(\mbox{lcoeff}_x(f)\right)=0 \mbox{ or }\varphi_a\left(\mbox{lcoeff}_x(g)\right)=0\}\\
{\mathcal B}_2^{(x,y)}(f,g)=\{a\in {\Bbb R}|\varphi_a\left(\mbox{lcoeff}_y(f)\right)=0 \mbox{ or }\varphi_a\left(\mbox{lcoeff}_y(g)\right)=0\}\\
{\mathcal B}_3^{(x,y)}(f,g)=\{a\in {\Bbb R}|\varphi_a\left(\Res_x(\bar{f},\bar{g})\right)\}=0\\
{\mathcal B}_4^{(x,y)}(f,g)=\{a\in {\Bbb R}|\varphi_a\left(\Res_y(\bar{f},\bar{g})\right)\}=0\\
{\mathcal B}(f,g)={\mathcal B}_1 \cup {\mathcal B}_2 \cup {\mathcal B}_3 \cup {\mathcal B}_4\\
\end{array}
\]

Then the following lemma holds. 

\begin{lemma} \label{Tech-lemma}
Let $f,g\in {\Bbb R}[x,y,\lambda]$. Then, for all $a\notin {\mathcal B}(f,g)$, it holds that $\gcd(\varphi_a(f),\varphi_a(g))=\varphi_a(\gcd(f,g))$.
\end{lemma}

{\bf Proof.} Writing $f=\bar{h}\cdot h$, $g=\bar{g} \cdot h$, it holds that
\[\gcd(\varphi_a(f),\varphi_a(g))=\gcd(\varphi_a(\bar{f}),\varphi_a(\bar{g}))\cdot \varphi_a(h)\]So, we have to prove that for
$a\notin {\mathcal B}(f,g)$,
$\gcd(\varphi_a(\bar{f}),\varphi_a(\bar{g}))=1$. Indeed, if this
equality does not hold then either
$\Res_x(\varphi_a(\bar{f}),\varphi_a(\bar{g}))=0$ or
$\Res_y(\varphi_a(\bar{f}),\varphi_a(\bar{g}))=0$. On the other
hand, since $a\notin {\mathcal B}(f,g)$ then the resultants behave
well under specializations (see Lemma 4.3.1 in \cite{Wi96}), and
hence $\Res_x(\varphi_a(\bar{f}),\varphi_a(\bar{g}))=0$ (resp.
$\Res_y(\varphi_a(\bar{f}),\varphi_a(\bar{g}))=0$) iff
$\varphi_a(\Res_x(\bar{f},\bar{g}))=0$ (resp.
$\varphi_a(\Res_y(\bar{f},\bar{g}))=0$). Nevertheless,
$\varphi_a(\Res_x(\bar{f},\bar{g}))=0$ (resp.
$\varphi_a(\Res_y(\bar{f},\bar{g}))=0$) cannot happen because
$a\notin {\mathcal B}(f,g)$. \qed

Now we fix the following notation:
\[\phi_{\lambda}(t)=(u(t,\lambda),v(t,\lambda))=
\left(\displaystyle{\frac{X_{11}(t,\lambda)}{X_{12}(t,\lambda)},\frac{X_{21}(t,\lambda)}{X_{22}(t,\lambda)}}\right),\]
and we introduce the following polynomials:
\[
\begin{array}{c}
G_1^{\phi}(t,s,\lambda)=X_{11}(t,\lambda)\cdot X_{12}(s,\lambda)-X_{12}(t,\lambda)\cdot X_{11}(s,\lambda)\\
G_2^{\phi}(t,s,\lambda)=X_{21}(t,\lambda)\cdot X_{22}(s,\lambda)-X_{22}(t,\lambda)\cdot X_{21}(s,\lambda)
\end{array}
\]
Moreover, we write $G_1^{\phi}=\bar{G}_1^{\phi}\cdot \gcd(G_1^{\phi},G_2^{\phi})$, $G_2^{\phi}=\bar{G}_2^{\phi}\cdot \gcd(G_1^{\phi},G_2^{\phi})$, and we denote:
\[
\begin{array}{c}
{\mathcal D}_1={\mathcal B}_1^{(t,s)}(G_1^{\phi},G_2^{\phi}),\mbox{ } {\mathcal D}_2={\mathcal B}_2^{(t,s)}(G_1^{\phi},G_2^{\phi})\\
{\mathcal D}_3={\mathcal B}_3^{(t,s)}(G_1^{\phi},G_2^{\phi}),\mbox{ } {\mathcal D}_4={\mathcal B}_4^{(t,s)}(G_1^{\phi},G_2^{\phi})\\
{\mathcal D}={\mathcal D}_1 \cup {\mathcal D}_2 \cup {\mathcal D}_3 \cup {\mathcal D}_4,\\
\end{array}
\]
(notice that ${\mathcal D}_1={\mathcal D}_2$, ${\mathcal D}_3={\mathcal D}_4$ because $G_1^{\phi},G_2^{\phi}$ are symmetric w.r.t. $t,s$).
Then, the following theorem holds. Here, we denote the evaluations of $G_1^{\phi},G_2^{\phi}$ at $\lambda=a$, as $\varphi_a(G_1^{\phi}), \varphi_a(G_2^{\phi})$, respectively.

\begin{theorem} \label{charact-properness}
The parametrization $\phi_{\lambda}(t)$ is proper for almost all
values of $\lambda$ iff $\deg_t(\gcd(G_1^{\phi},G_2^{\phi}))=1$,
i.e. iff $\gcd(G_1^{\phi},G_2^{\phi})=t-s$ (maybe multiplied by
some polynomial $\alpha(\lambda)\in {\Bbb R}[\lambda]$).
Furthermore, if this condition holds, then the only values of
$\lambda$ where $\phi_{\lambda}(t)$ may not be proper, are those
in ${\mathcal D}$.
\end{theorem}

{\bf Proof.} Let $a\in {\Bbb R}$. By Theorem 4.30 in \cite{SWPD},
$\phi_a(t)$ is proper iff
\[\deg_t\left(\gcd\left(\varphi_a(G_1^{\phi}),\varphi_a(G_2^{\phi})\right)\right)=1.\]However,
by Lemma \ref{Tech-lemma}, if $a\notin {\mathcal D}$ then
\[\gcd\left(\varphi_a(G_1^{\phi}),\varphi_a(G_2^{\phi})\right)=\varphi_a\left(\gcd\left(G_1^{\phi},G_2^{\phi}\right)\right)\]Hence,
if we prove that
\[\deg_t\left(\varphi_a\left(\gcd\left(G_1^{\phi},G_2^{\phi}\right)\right)
\right)=\deg_t\left( \gcd(G_1^{\phi},G_2^{\phi})\right)\]for
$a\notin {\mathcal D}$, then we have finished. Indeed, if this
equality does not hold, then the leading coefficient w.r.t. $t$ of
$\gcd(G_1^{\phi},G_2^{\phi})$ vanishes at $\lambda=a$. However, in
that case the leading coefficients of $G_1^{\phi},G_2^{\phi}$
w.r.t. $t$ both vanish at $\lambda=a$, and this cannot happen
because $a\notin {\mathcal D}$. Finally, notice that $t-s$ always
divides $\gcd(G_1^{\phi},G_2^{\phi})$; hence,
$\deg_t(\gcd(G_1^{\phi},G_2^{\phi}))=1$ iff
$\gcd(G_1^{\phi},G_2^{\phi})=t-s$ (perhaps multiplied by some
polynomial $\alpha(\lambda)$). \qed

Hence, Theorem \ref{charact-properness} gives us an algorithm for checking hypothesis ($H_1$). Moreover, if the
condition in Theorem \ref{charact-properness} holds one can determine the set ${\mathcal D}$, which contains the finitely
many values of the parameter where properness fails. If the condition does not hold, one can compute a reparametrization
 $\xi_{\lambda}(t)$ of the family, proper for almost all values of $\lambda$, by
 applying the reparametrization algorithm in Section 6.1.2, p.
 193 of \cite{SWPD}. We give more details in Subsection
 \ref{subsec-summary}.

\begin{remark} Observe that, if they exist, the $\lambda$-values making that the denominator of either $u$ or $v$ is identically $0$ belong to ${\mathcal D}$.
\end{remark}

\subsection{Checking hypothesis ($H_2$)} \label{subsec-dos}

Now let us consider hypothesis ($H_2$). For this purpose, we recall that the {\it degree of a rational function} (i.e. of a quotient of polynomials) is defined as the maximum of the degrees of the numerator and the denominator; furthermore, the {\it degree of a rational parametrization $\phi(t)=(\chi_1(t),\chi_2(t))$} is defined as the maximum of the degrees of $\chi_1(t),\chi_2(t)$ (which are rational functions). We also recall the following result from \cite{SWPD} (see Theorem 4.21 in \cite{SWPD}).

\begin{proposition} \label{prop-aux-degree}
If $\phi(t)=(\chi_1(t),\chi_2(t))$, where $\chi_1(t)$ is not identically $0$, is a proper rational parametrization of a curve and $f(x,y)=0$ is its implicit equation, then $\deg_t(\chi_1(t))=\deg_y(f)$
\end{proposition}

Moreover, we also need the following lemma (see for example Section 3 of \cite{LaloCompl}).

\begin{lemma} \label{lemm-aux-degree}
Let $f(x,y)$ implicitly define a plane curve ${\mathcal V}$, let $\mu\in {\Bbb R}$, and let $g_{\mu}(x,y)$ be the
implicit equation of the curve that is obtained from ${\mathcal V}$ by applying the linear transformation $\{x=X+\mu Y,y=Y\}$.
Then, for almost all values of $\mu$ it holds that $\deg_y(g_{\mu})=\deg(g_{\mu})$ (here, $\deg(g_{\mu})$ denotes the total degree of $g_{\mu}$).
\end{lemma}

Now the following result holds. This theorem provides a method for computing the degree of a rational curve just from its parametrization, without making use of its implicit equation (compare also with Theorem 6 and Theorem 7 in \cite{PD07}).

\begin{theorem} \label{th-degree-curve}
Let $\phi(t)=(\chi_1(t),\chi_2(t))$ be a proper rational parametrization of a curve ${\mathcal V}$, where $\chi_1(t)$ is not identically $0$; also
let $\chi_{\mu}(t)=\chi_1(t)-\mu \chi_2(t)$, with $\mu$ generic. Then, $\deg({\mathcal V})=\deg_t(\chi_{\mu}(t))$.
\end{theorem}

{\bf Proof.} Let ${\mathcal U}_{\mu}$ be the curve obtained from ${\mathcal V}$ by applying the linear transformation $\{x=X+\mu Y, y=Y\}$. Then, for all values of $\mu$ it holds that ${\mathcal U}_{\mu}$ is a rational curve properly parametrized by $\tilde{\phi}_{\mu}(t)=(\chi_{\mu}(t),\chi_2(t))$. Now let $g_{\mu}(x,y)$ be the implicit equation of ${\mathcal U}_{\mu}$. Then
by Lemma \ref{lemm-aux-degree}, for a generic $\mu$ it holds that $\deg_y(g_{\mu})=\deg(g_{\mu})$. Now since the degree of a curve is invariant by linear transformations, we have that $\deg({\mathcal V})=\deg(g_{\mu})$. Finally, since $\chi_1(t),\chi_2(t)$ are not both identically 0,
for a generic $\mu$ it holds that $\chi_{\mu}(t)$ is not identically $0$, either; moreover, $\tilde{\phi}_{\mu}(t)$ is proper for every $\mu\in {\Bbb R}$, and hence by Proposition \ref{prop-aux-degree} we have that
 $\deg_y(g_{\mu})=\deg_t(\chi_{\mu}(t))$. Therefore the statement follows. \qed

Theorem \ref{th-degree-curve}, together with Proposition \ref{prop-aux-degree}, provides the following corollary.

\begin{corollary} \label{corol-deg-curve}
Let $\phi(t)=(\chi_1(t),\chi_2(t))$, where $\chi_1(t)$ not identically $0$, be a proper rational parametrization of a curve
${\mathcal V}$, and let $f(x,y)=0$ be its implicit equation. Also, let $\chi_{\mu}(t)=\chi_1(t)-\mu \chi_2(t)$, with $\mu$
generic. Then, $\deg_t(\chi_1(t)))=\deg_t(\chi_{\mu}(t))$ iff $\deg_y(f)=\deg(f)$.
\end{corollary}

Then the following characterization of $(H_2)$ can be deduced.

\begin{corollary} \label{char-H-2}
$(H_2)$ is fulfilled iff,
for a generic $\mu$, it holds that
$\deg_t\left(u(t,\lambda)\right)=\deg_t\left(u(t,\lambda)-\mu v(t,\lambda)\right)$. Moreover, the above value provides the
degree of $F_{\lambda}$, as a polynomial in the variables $x,y$.
\end{corollary}

Also from \cite{LaloCompl}, one may see that if $(H_2)$ is not satisfied, almost all changes of coordinates of the type $\{x=X+\mu Y,y=Y\}$ set the surface properly. Furthermore, if $(H_2)$ is fulfilled, then the following proposition holds.

\begin{proposition} \label{prop-bad-values-lcoeff}
Assume that $(H_2)$ holds, and let $m=\deg_t(X_{11}(t,\lambda))$, $n=\deg_t(X_{12}(t,\lambda))$,
$a(\lambda)=\lcoeff_t(X_{11}(t,\lambda))$, $b(\lambda)=\lcoeff_t(X_{12}(t,\lambda))$. Also, let $\lambda_0$
be a real root of the leading coefficient of $F_{\lambda}(x,y)$ w.r.t. the variable $y$, so that:
(i) $X_{11}(t,\lambda_0)$ does not vanish identically; (ii) $\phi_{\lambda_0}(t)$ is proper. Then, the following statements are true:
\begin{itemize}
\item [(1)] If $m>n$, then $a(\lambda_0)=0$.
    \item [(2)] If $m=n$, then $a(\lambda_0)=b(\lambda_0)=0$.
    \item [(3)] If $m<n$, then $b(\lambda_0)=0$.
        \end{itemize}
        \end{proposition}

        {\bf Proof.} Since by hypothesis $(H_2)$ holds, if the leading coefficient of $F_{\lambda}(x,y)$ vanishes at $\lambda=\lambda_0$ then $\deg_y(F_{\lambda_0})<\deg_y(F_{\lambda})$. Also, because of the required conditions (i) and (ii), Proposition \ref{prop-aux-degree} holds and therefore $\deg_y(F_{\lambda_0})=\deg_t(\chi_1(t,\lambda_0))$. Moreover, from Theorem \ref{th-degree-curve} it holds that $\deg_y(F_{\lambda})=\deg_t(\chi_1(t,\lambda))$. So, we deduce that $\deg_t(\chi_1(t,\lambda_0))<\deg_t(\chi_1(t,\lambda))$. The rest follows from the definition of degree of a rational function. \qed

        Notice that if $X_{11}(t,\lambda_0)=0$, then $\lambda_0$ is an element of the set ${\mathcal D}$ in Theorem \ref{charact-properness}. Now Proposition \ref{prop-bad-values-lcoeff} provides the following corollary. Here, we keep the notation used in the above proposition.

        \begin{corollary} \label{lcoef}
        Assume that $(H_1)$ and $(H_2)$ are fulfilled. Then, the real roots of $\lcoeff_y(F)$ belong to the set consisting
        of the following elements: (i) the real elements of the set ${\mathcal D}$ in Theorem \ref{charact-properness};
        (ii) the real roots of: (a) $a(\lambda)$, if $m>n$; (b) $b(\lambda)$, if $m<n$; (c) $\gcd(a(\lambda),b(\lambda))$, if $m=n$.
        \end{corollary}

    \subsection{Normality}\label{subsec-normal}

    Given a parametrization $\psi(t)$ of a plane curve ${\mathcal E}$, one says that the parametrization is {\sf normal} if it is surjective, i.e. if for all $P_0\in {\mathcal E}$ there exists $t_0\in {\Bbb C}$ so that $\psi(t_0)=P_0$; notice that $t_0$ may be complex even though $P_0$ is real. This notion has been studied in \cite{AC07}, \cite{S02}, \cite{SWPD}. For our purposes, here we recall the following result (see Theorem 6.22 in \cite{SWPD}).

    \begin{theorem} \label{th-normal}
    Let $\phi(t)=\left(\displaystyle{\frac{X_{11}(t)}{X_{12}(t)},\frac{X_{21}(t)}{X_{22}(t)}}\right)$ be a parametrization of a plane curve ${\mathcal E}$, and let $n=\deg(X_{12})$, $s=\deg(X_{22})$. Also, let $b^{\star}=\coeff(X_{11},n)$ (i.e. the coefficient of degree $n$ in $X_{11}(t)$), $b=\coeff(X_{12},n)$, $d^{\star}=\coeff(X_{21},s)$, $d=\coeff(X_{22},s)$. Then,
    \begin{itemize}
    \item [(i)] If $m>n$ or $r>s$, then $\phi(t)$ is normal.
    \item [(ii)] If $m\leq n$ and $r\leq s$, then $\phi(t)$ is normal iff \[\deg_t\left(\gcd\left(b^{\star}X_{12}(t)-bX_{11}(t),d^{\star}X_{22}(t)-dX_{21}(t)\right)\right)\geq 1\]Moreover, if $\phi(t)$ is not normal, the only point that is not reached by the parametrization is the so-called ``critical point" $\left(\displaystyle{\frac{b^{\star}}{b},\frac{d^{\star}}{d}}\right)$, which is a point of ${\mathcal E}$.
        \end{itemize}
        \end{theorem}

        In our case, we deal with the family \[\phi_{\lambda}(t)=(u(t,\lambda),v(t,\lambda))=
\left(\displaystyle{\frac{X_{11}(t,\lambda)}{X_{12}(t,\lambda)},\frac{X_{21}(t,\lambda)}{X_{22}(t,\lambda)}}\right).\]Hence, from Theorem \ref{th-normal} we can derive the following result:

\begin{theorem} \label{th-normal-family}
Let $m=\deg_t(X_{11})$, $n=\deg_t(X_{12})$, $r=\deg_t(X_{21})$, $s=\deg_t(X_{22})$. Also, let $a(\lambda)=\coeff(X_{11}(t,\lambda),m)$, $b^{\star}(\lambda)=\coeff(X_{11}(t,\lambda),n)$, $b(\lambda)=\coeff(X_{12}(t,\lambda),n)$, $c(\lambda)=\coeff(X_{21}(t,\lambda),r)$, $d^{\star}(\lambda)=\coeff(X_{21}(t,\lambda),s)$, $d(\lambda)=\coeff(X_{22}(t,\lambda),s)$. Then, it holds that:
\begin{itemize}
\item [(i)] If $m>n$ and $r>s$, the values of $\lambda$ where $\phi_{\lambda}(t)$ may not be normal
satisfy $\gcd(a(\lambda),c(\lambda))=0$.
\item [(ii)] If $m>n$ and $r\leq s$, the values of $\lambda$ where $\phi_{\lambda}(t)$ may not be normal
satisfy $a(\lambda)=0$.
    \item [(iii)] If $r>s$ and $m\leq s$, the values of $\lambda$ where $\phi_{\lambda}(t)$ may not be normal
    satisfy $c(\lambda)=0$.
    \item [(iv)] If $m\leq n$, $r\leq s$, and
    \[\deg_t\left(\gcd\left(b^{\star}(\lambda)X_{12}(t)-b(\lambda)X_{11}(t),d^{\star}(\lambda)X_{22}(t)-d(\lambda)X_{21}(t)\right)\right)\geq 1,\]
    the values of $\lambda$ where $\phi_{\lambda}(t)$ may not be normal satisfy at least one of the following conditions (here, we denote $\eta(t,\lambda)=\lcoeff_t(b^{\star}X_{12}-bX_{11})$, $\nu(t,\lambda)=d^{\star}X_{22}-dX_{21}$, $\tilde{\eta}=\eta/\gcd(\eta,\nu)$, $\tilde{\nu}=\nu/\gcd(\eta,\nu)$):
        (a) $\lcoeff_t(\eta(t,\lambda))=0$; (b) $\lcoeff_t(\nu(t,\lambda))=0$; (c) $\Res_t(\tilde{\eta},\tilde{\nu})=0$.
        \item [(v)] If $n\leq m$, $r\leq s$, the only points of the surface $S$ that may not be reached by the parametrization, are the points of the space curve ${\mathcal C}_{crit}$ parametrized by  $\left(\displaystyle{\frac{b^{\star}(\lambda)}{b(\lambda)},\frac{d^{\star}(\lambda)}{d(\lambda)}},\lambda\right)$.
            \end{itemize}
            \end{theorem}

            {\bf Proof.} The statements (i), (ii), (iii) and (v) essentially follow from Theorem \ref{th-normal}. Statement (iv) follows from Theorem \ref{th-normal} and Lemma 4.26 in \cite{SWPD}. \qed


\subsection{Summary, and an example}\label{subsec-summary}

Here we provide the full algorithm for checking the hypotheses
$(H_1)$, $(H_2)$, required on the family ${\mathcal F}$, and we
illustrate it by means of an example. Besides checking $(H_1)$ and
$(H_2)$, the algorithm also provides a list of finitely many
`special" values of the parameter, which will be useful in the
next section. If $(H_1)$ does not hold, then one can use the
reparametrization algorithm in Section 6.1.2, p.
 193 of \cite{SWPD} in the following way
 (see \cite{SWPD} for further information): (1) let
$G^{\phi}=\gcd(G_1^{\phi},G_2^{\phi})$; then, choose
$(\alpha,\beta)\in {\Bbb R}^2$ so
 that $G^{\phi}(\alpha,\beta)$ is not identically 0 (those
 finitely many $\lambda$-values making $G^{\phi}(\alpha,\beta)=0$ are
 incorporated to the list of ``special" values); choose also $a,b,c,d\in {\Bbb R}$ so that $ad-bc\neq 0$. (2) Consider the
 rational function
 \[R_{\lambda}(t)=\displaystyle{\frac{aG^{\phi}(\alpha,t)+bG^{\phi}(\beta,t)}{cG^{\phi}(\alpha,t)+dG^{\phi}(\beta,t)}}\]Then,
 let $r=\deg(\phi_{\lambda})/\deg(R_{\lambda}(t))$. In general the value of $r$ has to
 be discussed upon the value of $\lambda$, but it will be constant
 except for finitely many $\lambda$-values, which again must
 be stored in the list of special values. (3) Introduce a generic rational
 parametrization ${\mathcal Q}(t)$ of degree $r$ (i.e. the generic
 value of $r$) with undetermined coefficients. From the equality
 $\phi_{\lambda}(t)={\mathcal Q}(R_{\lambda}(t))$ derive a linear system of equations
 in the undetermined coefficients, and by solving it determine
 ${\mathcal Q}(t)$ (in fact ${\mathcal Q}_{\lambda}(t)$). Notice that this linear system has $\lambda$ as a
 parameter, and therefore, again, certain (special) values of the parameter
 must also be computed (by discussing the system). In the end, the
 whole
 process yields a reparametrization (depending on $\lambda$), and a finite list of special
  values of $\lambda$. The correctness of
this process follows from Theorem 6.4, p. 191 of [18].

Also, if $(H_2)$ is not satisfied almost all changes of
coordinates of the type $\{X=x+\mu y,Y=y\}$ set the surface
properly; observe that this kind of transformation leaves the
$z$-coordinate invariant, and so the topology of the level curves
of the surface is not changed.

{\sf {\underline{Algorithm:} ({\tt Check})}} {\sf Given} a uniparametric family ${\mathcal F}$ of rational curves, defined by its parametric equations
\[\phi_{\lambda}(t)=(u(t,\lambda),v(t,\lambda))=
\left(\displaystyle{\frac{X_{11}(t,\lambda)}{X_{12}(t,\lambda)},\frac{X_{21}(t,\lambda)}{X_{22}(t,\lambda)}}\right),\]where $m=\deg_t(X_{11})$, $n=\deg_t(X_{12})$, $r=\deg_t(X_{21})$, $s=\deg_t(X_{22})$, and $a(\lambda)=\coeff(X_{11}(t,\lambda),m)$, $b^{\star}(\lambda)=\coeff(X_{11}(t,\lambda),n)$, $b(\lambda)=\coeff(X_{12}(t,\lambda),n)$, $c(\lambda)=\coeff(X_{21}(t,\lambda),r)$, $d^{\star}(\lambda)=\coeff(X_{21}(t,\lambda),s)$, $d(\lambda)=\coeff(X_{22}(t,\lambda),s)$, and
where $\lambda$ is a real parameter, the algorithm {\sf checks} hypotheses $(H_1)$ and $(H_2)$; moreover, if the hypotheses are satisfied, the algorithm also {\sf computes} a list ${\mathcal Spec}$ of finitely many ``special" values of $\lambda$.

\begin{itemize}
\item [1.] (Check hypothesis $(H_1)$)
\begin{itemize}
\item [1.1] Compute the polynomials $G_1^{\phi},G_2^{\phi}$ in Section \ref{subsec-uno}.
\item [1.2] If $\deg_t(\gcd(G_1^{\phi},G_2^{\phi}))=1$, then return {\tt $(H_1)$ holds}, otherwise return {\tt $(H_1)$ does not hold}.
\item [1.3] If $(H_1)$ does not hold, reparametrize the family.
Let ${\mathcal Spec}_0$ be the list of special values computed in
the process (if $(H_1)$ holds, ${\mathcal Spec}_0=\emptyset$).
\item [1.4] Compute the set ${\mathcal D}$ in Theorem \ref{charact-properness}, and let ${\mathcal Spec}_1:= {\mathcal D}$.
\end{itemize}
\item [2.] (Check hypothesis $(H_2)$)
\begin{itemize}
\item [2.1] Compute $\deg_t\left(u(t,\lambda)\right),\deg_t\left(u(t,\lambda)-\mu v(t,\lambda)\right)$. If both are equal, then return {\tt $(H_2)$ holds}, otherwise return {\tt $(H_2)$ does not hold}.
    \item [2.2] If $(H_2)$ does not hold, apply a change of coordinates $\{X=x+\mu y,Y=y\}$, and go back to 1.3.
    \item [2.3] If $(H_2)$ holds, then compute the real roots of: (a) $a(\lambda)$, if $m>n$; (b) $b(\lambda)$, if $m<n$; (c) $\gcd(a(\lambda),b(\lambda))$, if $m=n$. Let ${\mathcal Spec}_2$ be the set consisting of these values.
        \end{itemize}
        \item [3.] (Normality)
        \begin{itemize}
        \item [3.1] If $m>n$ and $r>s$, let ${\mathcal Spec}_3$ be the set of real roots of $\gcd(a(\lambda),c(\lambda))$.
        \item [3.2] If $m>n$ and $r\leq s$, let ${\mathcal Spec}_3$ be the set of real roots of $a(\lambda)$.
        \item [3.3] If $r>s$ and $m\leq n$, let ${\mathcal Spec}_3$ be the set of real roots of $c(\lambda)$.
        \item [3.4] If If $m\leq n$, $r\leq s$, then compute \[\delta=\deg_t\left(\gcd\left(b^{\star}(\lambda)X_{12}(t)-b(\lambda)X_{11}(t),d^{\star}(\lambda)X_{22}(t)-d(\lambda)X_{21}(t)\right)\right)\]
            \begin{itemize}
            \item [3.4.1] If $\delta\geq 1$, then let $\eta(t,\lambda):=\lcoeff_t(b^{\star}X_{12}-bX_{11})$, $\nu(t,\lambda):=d^{\star}X_{22}-dX_{21}$, $\tilde{\eta}:=\eta/\gcd(\eta,\nu)$, $\tilde{\nu}:=\nu/\gcd(\eta,\nu)$, and let ${\mathcal Spec}_3$ be the set of real roots of $\lcoeff_t(\eta(t,\lambda))$, $\lcoeff_t(\nu(t,\lambda))$,  $\Res_t(\tilde{\eta},\tilde{\nu})=0$.
                \item [3.4.2] If $\delta<1$ then determine the parametrization \[\left(\displaystyle{\frac{b^{\star}(\lambda)}{b(\lambda)},\frac{d^{\star}(\lambda)}{d(\lambda)}},\lambda\right)\] of the space curve ${\mathcal C}_{crit}$. Let ${\mathcal Spec}_3:=\emptyset$
                    \end{itemize}
                    \end{itemize}
            \item [4.] {\bf Special values:} ${\mathcal Spec}:={\mathcal Spec}_0\cup {\mathcal Spec}_1 \cup {\mathcal Spec}_2 \cup {\mathcal Spec}_3$ \end{itemize}

The above algorithm is illustrated by the next example. Here, we consider the parametric equation of the offset family to a cardioid.
By using the results in \cite{ASS96}, one may check that this offset family is rational; moreover, the parametrization used in the
example is taken from \cite{Juani-thesis}.

\begin{example} \label{prep-off-cardiod}
Let ${\mathcal O}_d({\mathcal C})$ be the offset family to the
cardioid $(x^2+4y+y^2)^2-16(x^2+y^2)=0$. Then, a parametrization
of this family is $\phi_d(t)=(u(t,d),v(t,d))$, where:
\[
\begin{array}{rcl}
u(t,d) &=& \displaystyle{\frac{3456t^5-31104t^3+dt^8-126dt^6+10206dt^2-6561d}{486t^4+36t^6+2916t^2+t^8+6561}}\\
v(t,d) &= & \displaystyle{\frac{(-18)t(864t^3-16t^5-1296t+dt^6-21dt^4-189dt^2+729d)}{486t^4+36t^6+2916t^2+t^8+6561}}
\end{array}
\]

Here, the parameter $d$ denotes the offsetting distance. So, let us apply the algorithm {\tt Check}. In step (1), we analyze $(H_1)$. For this
purpose, we compute $G_1^{\phi},G_2^{\phi}$ (step 1.1):

$G_1^{\phi}=(3456t^5-31104t^3+dt^8-126dt^6+10206dt^2-6561d)(486s^4+36s^6+2916s^2+s^8+6561)-(3456s^5-31104s^3+ds^8-126ds^6+10206ds^2-6561d)(486t^4+36t^6+2916t^2+t^8+6561)$
\newline

$G_2^{\phi}=-18t(864t^3-16t^5-1296t+dt^6-21dt^4-189dt^2+729d)(486s^4+36s^6+2916s^2+s^8+6561)+18s(864s^3-16s^5-1296s+ds^6-21ds^4-189ds^2+729d)(486t^4+36t^6+2916t^2+t^8+6561)$

Then, we check that $\gcd(G_1^{\phi},G_2^{\phi})=t-s$, and
therefore that $\deg_t(\gcd(G_1^{\phi},G_2^{\phi}))=1$ (step
(1.2)); hence, $(H_1)$ holds, ${\mathcal Spec}_0:=\emptyset$, and
we go to step (1.4). Here, we compute the set ${\mathcal D}$
containing the $d$-values where $\phi_d(t)$ may not be proper; for
this purpose, we compute the real roots of: (i)
$\mbox{Content}_s(\lcoeff(G_1^{\phi},t))$; (ii)
$\mbox{Content}_s(\lcoeff(G_2^{\phi},t))$; (iii)
$\mbox{Content}_t\left(\Res_t(\bar{G}_1^{\phi},
\bar{G}_2^{\phi})\right)$. In this case, we get that
$\mbox{Content}_s(\lcoeff(G_1^{\phi},t))=54$,
$\mbox{Content}_s(\lcoeff(G_2^{\phi},t))=18$, and
$\Res_t(\bar{G}_1^{\phi}, \bar{G}_2^{\phi})=C\cdot
d(t^2+9)^{44}(729d-1053dt^2-117dt^4+dt^6+3456t^3)(-32t+dt^2+9d)^2$
(with $C\in {\Bbb N}$). Hence, it holds that ${\mathcal D}=\{0\}$
and therefore \[{\mathcal Spec}_1:=\{0\}.\]Now in step (2), we
check $(H_2)$. For this purpose, in step (2.1) we compute
$\deg_t(u(t,d))=8$ and $\deg_t(u(t,d)-\mu v(t,d)=8$; since both of
them coincide, then $(H_2)$ holds, and we go to step (2.3). Since
$m=n=8$, we compute $a(d)=d$, $b(d)=1$, and we determine
$\gcd(a(d),b(d))$, which is 1; hence,  \[{\mathcal Spec}_2:=
\emptyset .\] Finally, in step (3) we consider normality
questions. We have that $r=7,s=8$ and therefore $r<s$. Since
$m=n$, we go to (3.4) and we compute
\[
\begin{array}{rcl}
\eta(t,d) & = & d(486t^4+36t^6+2916t^2+t^8+6561)-3456t^5+31104t^3-dt^8+126dt^6\\
&&-10206dt^2+6561d\\
\nu(t,d) &= & 18dt(864t^3-16t^5-1296t+dt^6-21dt^4-189dt^2+729d)\\
\delta & = & \deg_t(\gcd(\eta,\nu))
\end{array}
\]
We get $\delta=0$; hence, we go to step (3.4.2) and we obtain a
space curve of possibly non-reached points, namely
\[{\mathcal C}_{crit}=\left(d,0,d\right)\]
Finally, we get that ${\mathcal Spec}:=\{0\}$.
\end{example}

\section{The algorithm for the parametric case.} \label{sec-param-initial}

Based on the observations made in Subsection \ref{important-obs},
in this section we present an algorithm for computing a critical
set of $S$, under the assumption that the hypotheses requested in
Subsection \ref{subsec-statement} are satisfied; the algorithm
does not compute or make use of the implicit equation of $F$. More
precisely, in Subsection \ref{important-obs} it is observed that
one can compute a critical set of $S$ by determining certain
$z$-values referred to as {\sf $(1)$-values}, {\sf $(2)$-values}
and {\sf $(3)$-values}, respectively. These values have different
geometric meanings, and are related to notable points and lines of
the curve ${\mathcal M}$ defined in the $xz$-plane by the
polynomial $M(x,z)=\sqrt{D_y(F)}$, where $F$ is the implicit
equation of $S$. On the other hand, also in Subsection
\ref{important-obs} it is shown that ${\mathcal M}$ can be written
as the union of the projection onto the $xz$-plane of the space
curve ${\mathcal C}=V(F,F_y)$, and the curve (in the $xz$-plane)
defined by $\lcoeff_y(F)$. Now, the real roots of $\lcoeff_y(F)$
(or, more precisely, a finite set containing them) can be
determined by means of Corollary \ref{lcoef}, without explicitly
computing $F$. Hence, in this section we focus on the computation
of the remaining (1), (2) and (3)-values, which are related to
${\mathcal C}$; for this purpose, and since we do not want to
compute or make use of $F$, the idea is to work with a
``parametric description" of ${\mathcal C}$.

In order to provide equations for ${\mathcal C}$, one may see (recall the first paragraph in Subsection \ref{important-obs})
that this curve consists of the following points:

\begin{itemize}
\item points of $S$, reached by the parametrization, such that the normal vector to $S$ is either $\vec{0}$ (in which case the point is a singularity of $S$), or parallel to the $xz$ plane.
    \item self-intersections of $S$, reached by the parametrization.
    \item points of $S$ with some of the above geometric properties, but not reached by the parametrization.
    \end{itemize}

    In the sequel we will refer to these sets as {\sf first, second} and {\sf third} sets, respectively. Notice that some of these sets may be
    empty, and that they are not necessarily disjoint. So, we start with the first one.
    By performing easy computations with the parametrization $(u(t,\lambda),v(t,\lambda),\lambda)$ of $S$, one may see that the expression of the normal vector to $S$ is:
    \[\vec{N}=v_t\vec{i}-u_t\vec{j}+(u_tv_{\lambda}-u_{\lambda}v_t)\vec{k}\]Hence, we define ${\mathcal C}_1$ as the following subset of points $(x,y,z)\in {\Bbb C}^3$:
    \[
{\mathcal C}_1\equiv \left  \{ \begin{array}{l}
x=u(t,\lambda)\\
y=v(t,\lambda)\\
z=\lambda\\
h(t,\lambda)=0\\
X_{12}(t,z)\cdot X_{22}(t,z)\neq 0\\
\end{array} \right.,
\]
    where $h(t,\lambda)$ is the square-free part of the numerator of $u_t(t,\lambda)$. One may see that ${\mathcal C}_1$ contains
    the first set. Then let us consider the second set (i.e. self-intersections of $S$). For this purpose, we impose
    \[u(t,\lambda)=u(s,\lambda), v(t,\lambda)=v(s,\lambda), \mbox{ with }t\neq s\]Hence, we find again the
    polynomials $G_1^{\phi},G_2^{\phi}$ introduced in Subsection \ref{subsec-uno}. We recall from
    Subsection \ref{subsec-uno} the
    notation $\bar{G}_1^{\phi},\bar{G}_2^{\phi}$ for the
    polynomials obtained by removing the common factor $t-s$ in
    $G_1^{\phi},G_2^{\phi}$. Furthermore, we write $j(t,\lambda)=\sqrt{\Res_s(\bar{G}_1^{\phi},\bar{G}_2^{\phi})}$. Since
by hypothesis the parametrization
$\phi_{\lambda}(t)=(u(t,\lambda),v(t,\lambda),\lambda)$ is proper
for almost all values of $\lambda$, then $j(t,\lambda)$ cannot be
identically 0 (see Theorem 4.30 in \cite{SWPD}). Therefore, we
define ${\mathcal C}_2$ as the following subset of points
$(x,y,z)\in {\Bbb C}^3$:
\[
 {\mathcal C}_2\equiv \left\{ \begin{array}{l}
x=u(t,\lambda)\\
y=v(t,\lambda)\\
z=\lambda\\
j(t,\lambda)=0\\
X_{12}(t,z)\cdot X_{22}(t,z)\neq 0\\
\end{array} \right.
\]
One may see that this set contains the self-intersections of $S$
reached by the parametrization. Finally, the third set of points
has been studied in Subsection \ref{subsec-normal}. In this sense,
by Theorem \ref{th-normal-family} we have basically two
possibilities: in the first case (see statements (i), (ii), (iii),
(iv) in Theorem \ref{th-normal-family}), there exists a finite set
${\mathcal N}$ containing the $\lambda$-values so that
$\phi_{\lambda}(t)$ may not be normal; in that situation, we just
add ${\mathcal N}$ to the rest of elements of the critical set,
and therefore no special difficulty arises. In the second case
(statement (v) of Theorem \ref{th-normal-family}), it may happen
that there are infinitely many $\lambda$ values so that
$\phi_{\lambda}(t)$ is not normal; thus, the non-reached points of
$S$ are among the points of the space curve ${\mathcal C}_{crit}$
parametrized by
$\left(\displaystyle{\frac{b^{\star}(\lambda)}{b(\lambda)},\frac{d^{\star}(\lambda)}{d(\lambda)}},\lambda\right)$.
Hence, in the sequel we will assume that we are in this more
complicated situation.

\subsection{Computation of a critical set}\label{comp-crit-set}

Since the roots of $\lcoeff_y(F)$ can de determined from Corollary
\ref{lcoef}, we focus on the remaining elements of the critical
set. For this purpose, we need to analyze the projection of
${\mathcal C}$ onto the $xz$-plane. From the above reasonings, we
can write
\[{\mathcal C}\subset {\mathcal C}_1 \cup {\mathcal C}_2 \cup {\mathcal C}_{crit}\]

So, $\pi_{xz}({\mathcal C})\subset \pi_{xz}({\mathcal C}_1) \cup \pi_{xz}({\mathcal C}_2) \cup \pi_{xz}({\mathcal C}_{crit})$. Now we have that
\[ \pi_{xz}({\mathcal C}_1) \equiv \left\{ \begin{array}{l}
x=u(t,z)\\
h(t,z)=0\\
X_{12}(t,z)\cdot X_{22}(t,z)\neq 0\\
\end{array} \right.
\]
Let $f(x,t,z)$ be the numerator of $x-u(t,z)$. Also, we write
$h=\tilde{h}\cdot \gcd(h,X_{12}\cdot X_{22})$. Then, we denote the
curve defined by $m^{(1)}(x,z)=\sqrt{\Res_t(f,\tilde{h})}$ as
${\mathcal M}_1$; notice that $m^{(1)}(x,z)$ cannot be identically
0 because by hypothesis the function $u(t,\lambda)$ is given in
reduced form (i.e. its numerator and denominator share no common
factor). Furthermore, one may see that $\pi_{xz}({\mathcal C}_1)$
is included in ${\mathcal M}_1$. Similarly,
\[ \pi_{xz}({\mathcal C}_2) \equiv \left\{ \begin{array}{l}
x=u(t,z)\\
j(t,z)=0\\
X_{12}(t,z)\cdot X_{22}(t,z)\neq 0\\
\end{array} \right.
\]
In this case, we write $j=\tilde{j}\cdot \gcd(h,X_{12}\cdot
X_{22})$, we denote $m^{(2)}(x,z)=\sqrt{\Res_t(f,\tilde{j})}$
(which cannot be identically 0 for the same reason than
$m^{(1)}(x,z)$), and we represent the curve defined by $m^{(2)}$
as ${\mathcal M}_2$. One may observe that $\pi_{xz}({\mathcal
C}_2)$ is included in ${\mathcal M}_2$. Finally,
$\pi_{xz}({\mathcal C}_{crit})$ is the parametric curve defined by
$\left(\displaystyle{\frac{b^{\star}(\lambda)}{b(\lambda)}},\lambda\right)$.

Now, the $z$-coordinates of the singularities and of the points with tangent parallel to the $x$-axis of $
{\mathcal M}_1$ (resp. ${\mathcal M}_2$), and also the values $z_a$ so that $z-z_a$ is a horizontal asymptote
of ${\mathcal M}_1$ (resp. ${\mathcal M}_2$), correspond to real roots of $\Res_x(m^{(1)}, m_x^{(1)})$ (resp. $\Res_x(m^{(2)}, m_x^{(2)})$). On the other hand, since $\pi_{xz}({\mathcal C}_{crit})$ is parametrized by $\left(\displaystyle{\frac{b^{\star}(\lambda)}{b(\lambda)}},\lambda\right)$, it has no point with tangent parallel to the $x$-axis, no singularity, and no component parallel to the $x$-axis. However, it may have horizontal asymptotes, corresponding to the roots of $b(\lambda)$.

Finally, in order to compute a critical set we also need to
determine the intersections between ${\mathcal M}_1$ and
${\mathcal M}_2$, ${\mathcal M}_1$ and $\pi_{xz}({\mathcal
C}_{crit})$, ${\mathcal M}_2$ and $\pi_{xz}({\mathcal C}_{crit})$,
respectively. In the first case, let
\[\overline{m^{(1)}}=m^{(1)}/\gcd(m^{(1)},m^{(2)}), \overline{m^{(2)}}=m^{(2)}/\gcd(m^{(1)},m^{(2)});\]
then, the
$z$-coordinates of the intersections between ${\mathcal M}_1$ and
${\mathcal M}_2$ are contained in the set of roots of
$\Res_x(\overline{m^{(1)}},\overline{m^{(2)}})$. In the second
(resp. the third) case, we simply compute the roots of the
numerator $M^{(1)}(z)$ (resp. $M^{(2)}(z)$) where $M^{(1)}(z)$
(resp. $M^{(2)}(z)$) is defined as the result of substituting
$x:=\displaystyle{\frac{b^{\star}(\lambda)}{b(\lambda)}}$,
$z:=\lambda$ in  $m^{(1)}(x,z)$ (resp. $m^{(2)}(x,z)$), whenever
this substitution does not yield 0, and is defined as $1$ in other
case.

So, we can derive the following algorithm for computing a critical set of ${\mathcal F}$. Here, we will use the subset ${\mathcal Spec}$
computed by the algorithm {\tt Check}, that we developed in Section \ref{sec-checking}.

{\sf {\underline{Algorithm:} ({\tt Critical})}} {\sf Given} a uniparametric family ${\mathcal F}$ of rational curves depending on a parameter $\lambda$, defined by its parametric equations
\[\phi_{\lambda}(t)=(u(t,\lambda),v(t,\lambda))=
\left(\displaystyle{\frac{X_{11}(t,\lambda)}{X_{12}(t,\lambda)},\frac{X_{21}(t,\lambda)}{X_{22}(t,\lambda)}}\right),\]fulfilling hypotheses $(H_1),(H_2),(H_3)$, where $m=\deg_t(X_{11})$, $n=\deg_t(X_{12})$, $r=\deg_t(X_{21})$, $s=\deg_t(X_{22})$, and $b^{\star}(\lambda)=\coeff(X_{11}(t,\lambda),n)$, $b(\lambda)=\coeff(X_{12}(t,\lambda),n)$, $d^{\star}(\lambda)=\coeff(X_{21}(t,\lambda),s)$, $d(\lambda)=\coeff(X_{22}(t,\lambda),s)$,
the algorithm {\sf computes} a critical set ${\mathcal A}$ of the family.

\begin{itemize}
\item [(1)] Compute the set ${\mathcal A}_1$ consisting of the real roots of the following polynomials:
\begin{itemize}
\item $\Res_x(m^{(1)}, m_x^{(1)})$
\item $\Res_x(m^{(2)}, m_x^{(2)})$.
\item $\Res_x(\overline{m^{(1)}},\overline{m^{(2)}})$.
\end{itemize}
\item [(2)] If $m>n$, or $r>s$, or $m\leq n,r\leq s$ but \[\deg_t(\gcd(b^{\star}(\lambda)X_{12}(t)-b(\lambda)X_{11}(t),c(\lambda)X_{22}(t)-d(\lambda)X_{21}(t)))\geq1,\]then ${\mathcal A}_2=\emptyset$.
    Otherwise, let ${\mathcal A}_2$ be the set consisting of the real roots of the following polynomials:
    \begin{itemize}
\item $b(\lambda)$.
\item $M^{(1)}(z)$
\item $M^{(2)}(z)$
\end{itemize}
\item [(3)] Let ${\mathcal A}={\mathcal Spec}\cup {\mathcal A}_1\cup {\mathcal A}_2$. {\tt Return} ${\mathcal A}$.
    \end{itemize}

\subsection{Correctness of the Algorithm} \label{subsec-correct}

The aim of this subsection is to prove that the algorithm {\tt
Critical}, provided in the above subsection, is correct, i.e. that
the set ${\mathcal A}$ determined by the algorithm is a critical
set of $S$. The necessity of this proof is due to the fact that in
{\tt Critical} we are working not really with $\pi_{xz}({\mathcal
C})$, but with the curve ${\mathcal M}^{\star}={\mathcal M}_1\cup
{\mathcal M}_2 \cup \pi_{xz}({\mathcal C}_{crit})$, which in
general may be bigger than $\pi_{xz}({\mathcal C})$. Hence, we
want to ensure that no critical value (i.e. no $\lambda$-value
where
 the topology of the family changes) has been missed when passing from $\pi_{xz}({\mathcal C})$ to ${\mathcal
 M}^{\star}$. This is done in the following theorem.



\begin{theorem} \label{th-final}
The algorithm {\tt Critical} is correct.
\end{theorem}

{\bf Proof.} Since $\pi_{xz}({\mathcal C})\subset {\mathcal
M}^{\star}$, from Chapter 3, Section 2.5, Exercise 5 of
 \cite{walker} it follows that the singularities of $\pi_{xz}({\mathcal
 C})$ are also singularities of ${\mathcal
M}^{\star}$, and hence their $z$-coordinates are found by the
algorithm. Now, let $Q\in \pi_{xz}({\mathcal C})$ so that the
tangent line $\ell$ to $\pi_{xz}({\mathcal C})$ at $Q$ is parallel
to the $x$-axis. If $Q$ is a regular point of ${\mathcal
M}^{\star}$, since $\pi_{xz}({\mathcal C})\subset {\mathcal
M}^{\star}$ we have that $\ell$ is also the tangent line to
${\mathcal M}^{\star}$ at $Q$; therefore $Q$ is a point of
${\mathcal M}^{\star}$ with tangent parallel to the $x$-axis, and
hence its $z$-coordinate is found by the algorithm. Otherwise, $Q$
is a singular point of ${\mathcal M}^{\star}$ and therefore its
$z$-coordinate is also determined by the algorithm. On the other
hand,
 asymptotic branches of $\pi_{xz}({\mathcal C})$ and components lying in planes normal to the $z$-axis, keep their nature
 when passing to ${\mathcal M}^{\star}$. Finally, since the real roots of $\lcoeff_y(F)$ are included in ${\mathcal Spec}$,
 we conclude that all the (1), (2) and (3)-values are computed by {\tt Critical}, and the result follows. \qed

\begin{remark} \label{optimal}
Notice that we are stating that the set computed by the algorithm
{\tt Critical} is a critical set, i.e. that it contains all the
$\lambda$-values where the topology type of the family may change.
But we are not stating that the set that we are computing is
optimal. So, the output of the algorithm may include additional
$z$-values where the topology of the family does not change. This
drawback was already present in the algorithm of \cite{JGRS},
which was not optimal, either. However, our algorithm may yield
critical sets which are bigger than those in \cite{JGRS}. Since
affine transformations preserve the topology of curves, one can
recognize some superfluous values in a computed critical set in
the following way: (1) apply a random affine transformation
$\{X=x+\mu y, Y=y\}$ to the family; (2) compute a critical set of
the new family; (3) discard those values of the original critical
set that do not belong to the new critical set.
\end{remark}


\subsection{Experimentation and Results}\label{exp-res}

The algorithm {\tt Critical} has been implemented in Maple, and
tested with several examples. In this subsection, first we
continue the analysis of the topology types in the offset family
to the cardioid, started in Example \ref{prep-off-cardiod}. Then
we provide a simple example illustrating the non-optimality of the
algorithm, as mentioned in Remark \ref{optimal}. Finally we
present a table comparing timings between the algorithm for the
implicit case, deducible from \cite{JGRS}, and our algorithm.

{\bf Example \ref{prep-off-cardiod} (continued):} {\it By applying the
algorithm {\tt Critical}, one determines the following critical set:
\[{\mathcal A}=\{ -16/3, -\alpha, -3\sqrt{3}, -8\sqrt{3}/3, -3\sqrt{3}/2,0,
  3\sqrt{3}/2, 8\sqrt{3}/3, 3\sqrt{3}, \alpha, 16/3  \}\]
(which coincides with the output of the implicit algorithm), where
$\alpha$ is the real root of $729\lambda
^5-1215\lambda^4+702\lambda^3-18\lambda^2+13\lambda-27$. The total
amount of time required for this computation was 1.5 seconds (the
cost of checking the hypotheses is included). From this critical
set, one may deduce that there are at most 19 different topology
types in the family. However, by applying a random linear
transformation as suggested in Remark \ref{optimal}, one can
compute a reduced critical set, namely
\[\{ -16/3, -3\sqrt{3},  0,  3\sqrt{3}, 16/3  \}\]
In this case, because of the properties of offset curves, one has
that for $d=0$ one gets the original curve, i.e. a cardioid, and
that for $d_0$ and $-d_0$, the shape is the same. So, in our
analysis
 we have just considered positive values of $d$. In
Figure 2 one may find the different shapes arising in the family,
and the intervals corresponding to each of them. In the first row
(at the top of the figure), we display the pictures (i), (ii),
(iii) corresponding to the distances $d=1$, $d=3$, $d=22/5$,
respectively, all of them belonging to the interval $(0,
3\sqrt{3}/2)$ and therefore sharing the same topology type (we
have plotted the three pictures so as to clearly see the evolution
of the family, as $d$ is increased). In the second row, from left
to right we have the picture (iv) corresponding to $d=3\sqrt{3}$,
the picture (v) corresponding to a distance $d\in (3\sqrt{3},
16/3)$, and (vi), that corresponds to $d=16/3$. In the third row,
the shape (vii) corresponding to $d>16/3$ is shown. Also, in each
figure we have included the plotting of the original cardioid. One
may see that the offsets exhibit two cusps for $d<3\sqrt{3}$, and
a loop for $d\geq 16/3$. However, the topologies of (iv) and (v)
are not completely clear, since the picture does not show well
enough the behavior next to the singularity with negative
$y$-coordinate. If one enlarges the part of the curve next to this
singularity, one obtains the pictures in Figure 3. Here we have
plotted a detail of (iv) (left), of (v) (middle), and of (vi)
(right). So, in (iv) there is a non-ordinary singularity; in (v)
there is not one, but two singularities, corresponding to two
self-intersections of the curve, giving rise to two different
loops; in (vi), the topology changes so that the curve has only
one loop (the origin, in this case, is a singular point).}

\begin{figure}[ht]
\begin{center}
\centerline{$\begin{array}{ccc}
\psfig{figure=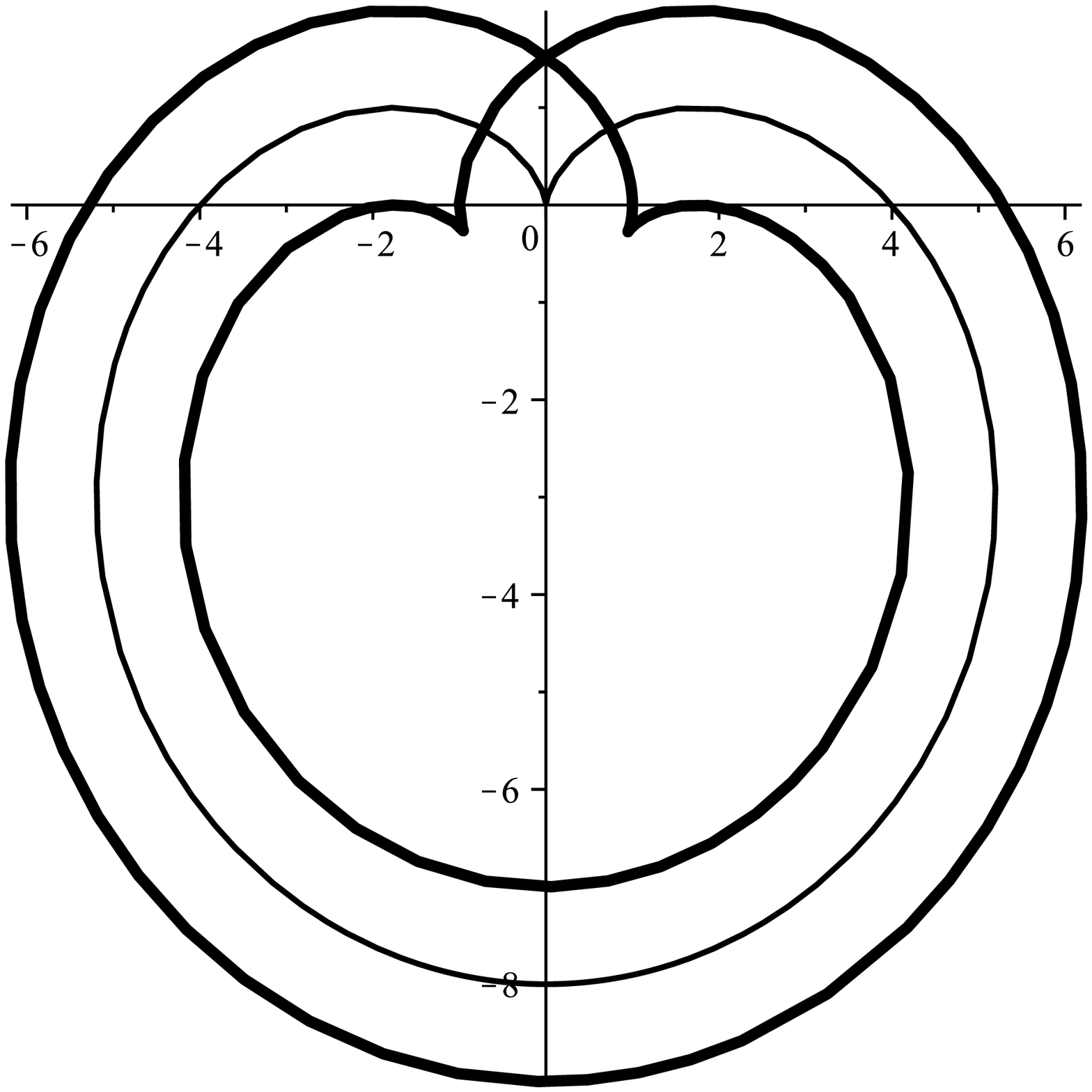,width=5cm,height=5cm} &
\psfig{figure=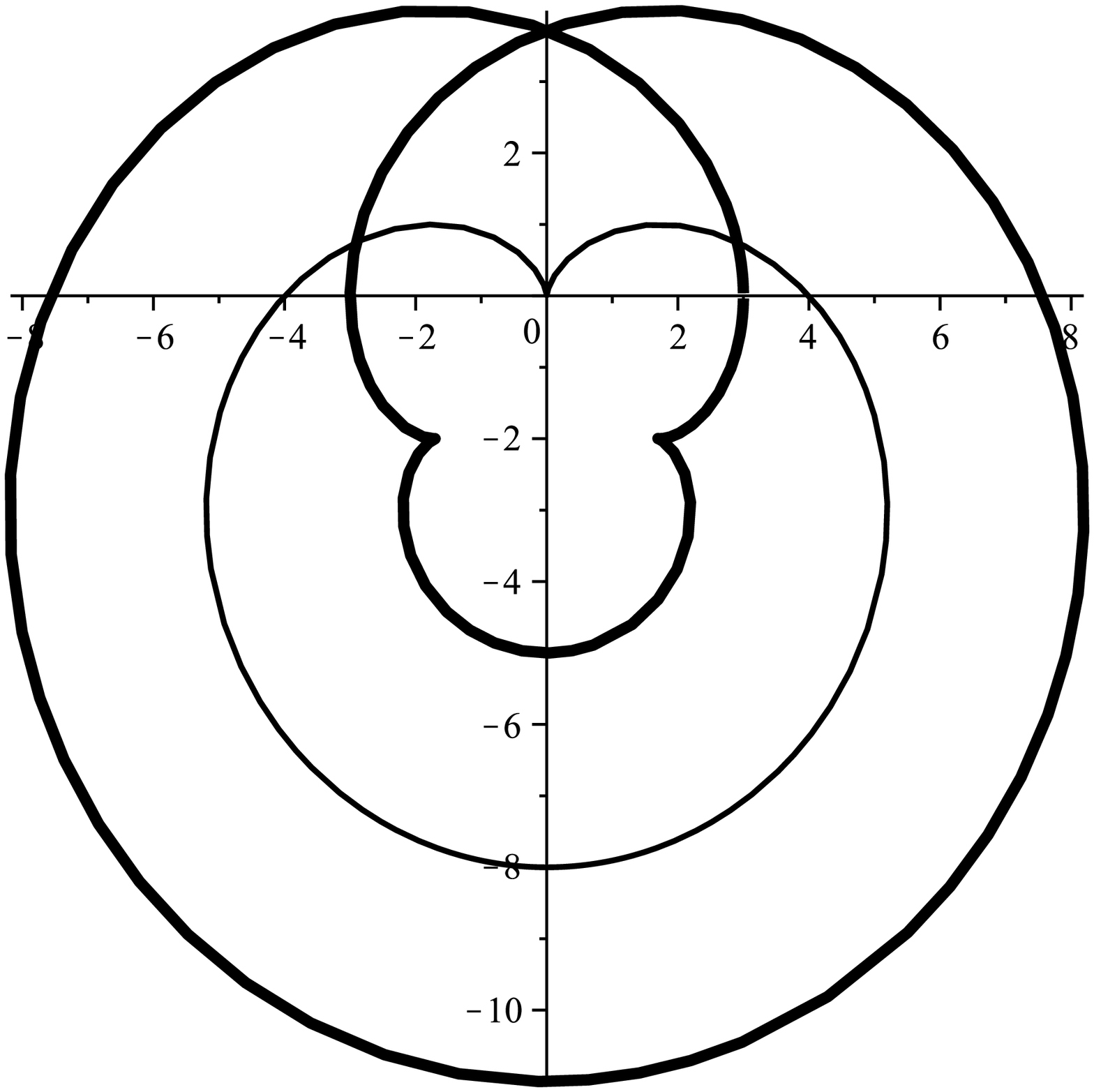,width=5cm,height=5cm} &
\psfig{figure=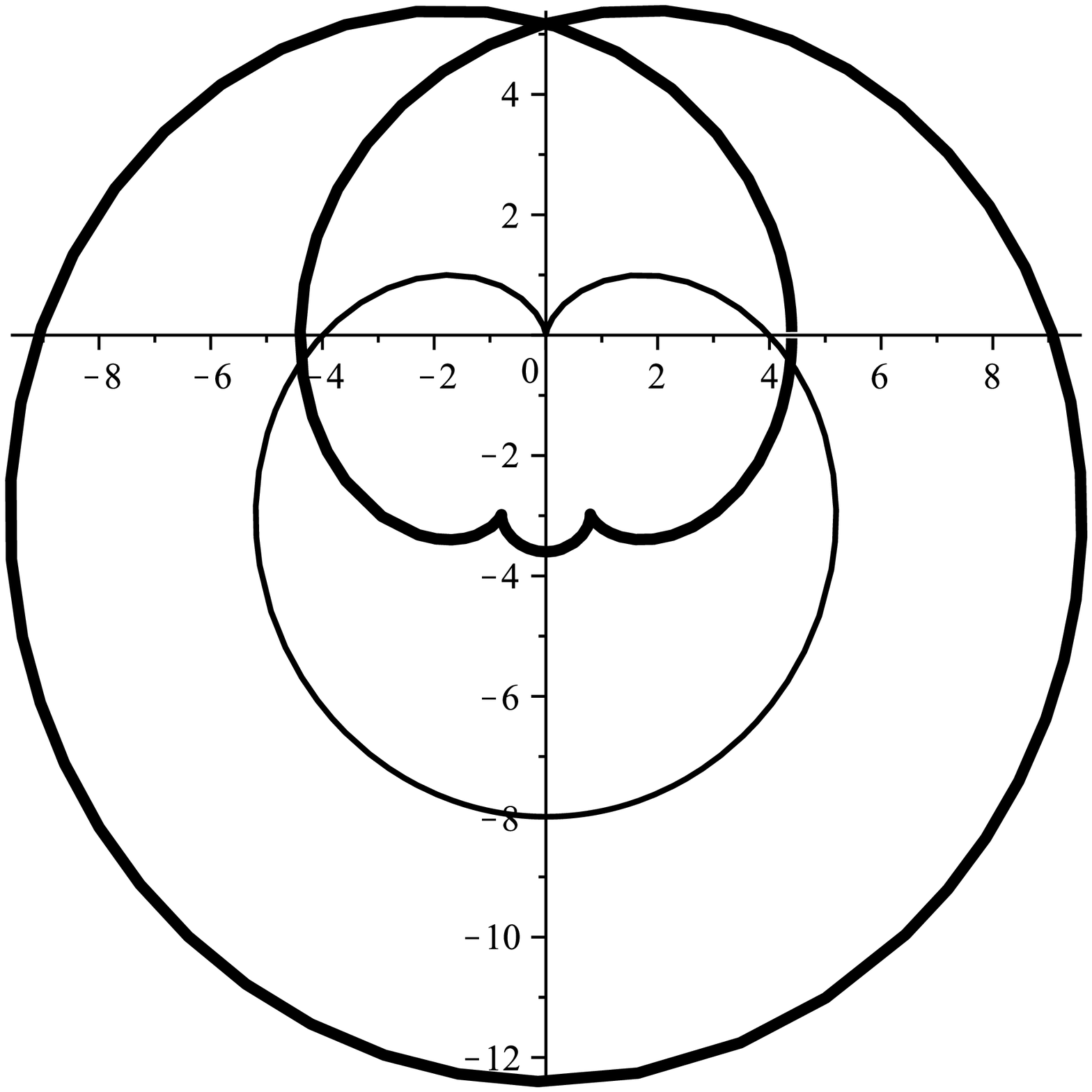,width=5cm,height=5cm} \\
\psfig{figure=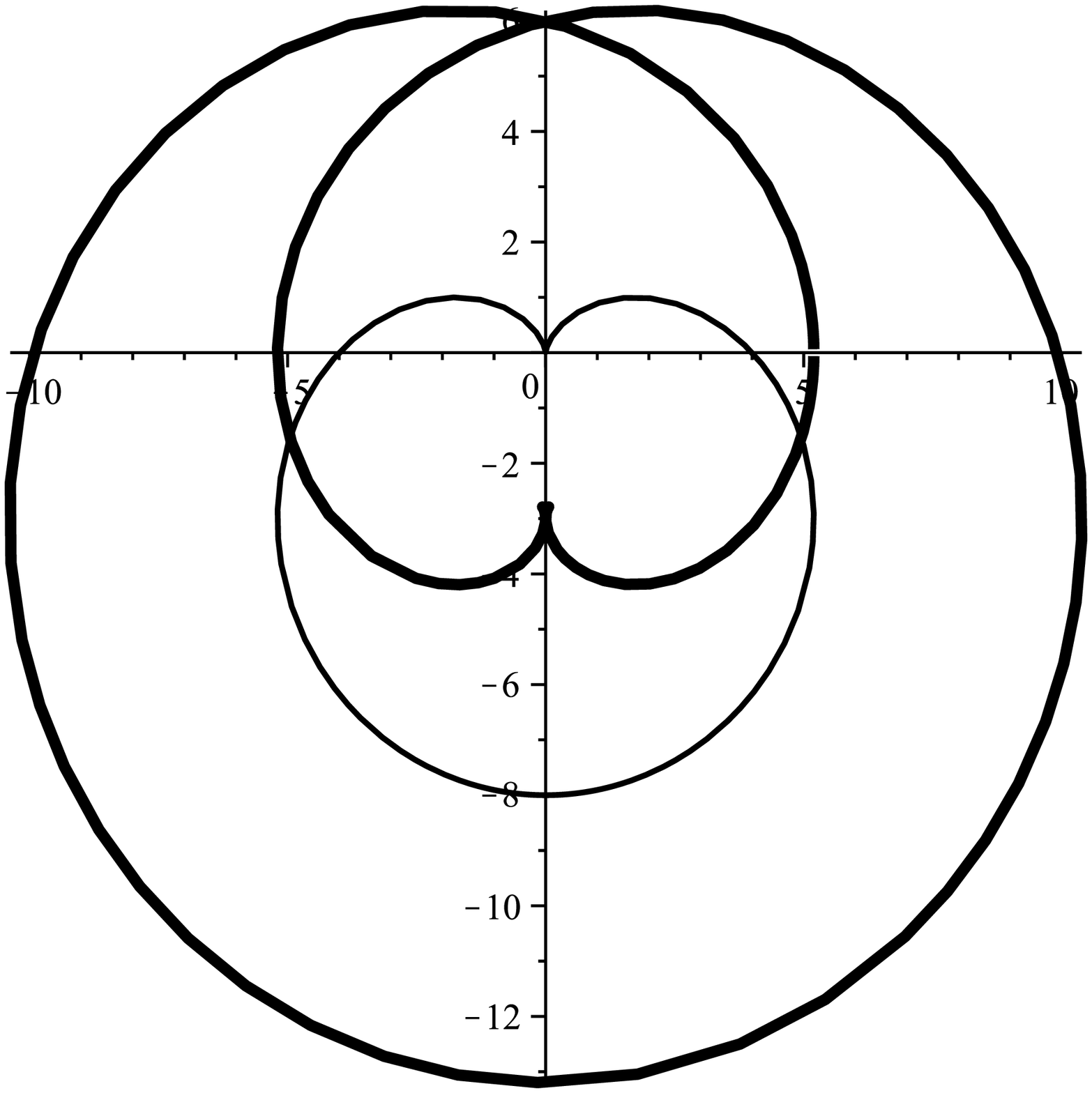,width=5cm,height=5cm} &
\psfig{figure=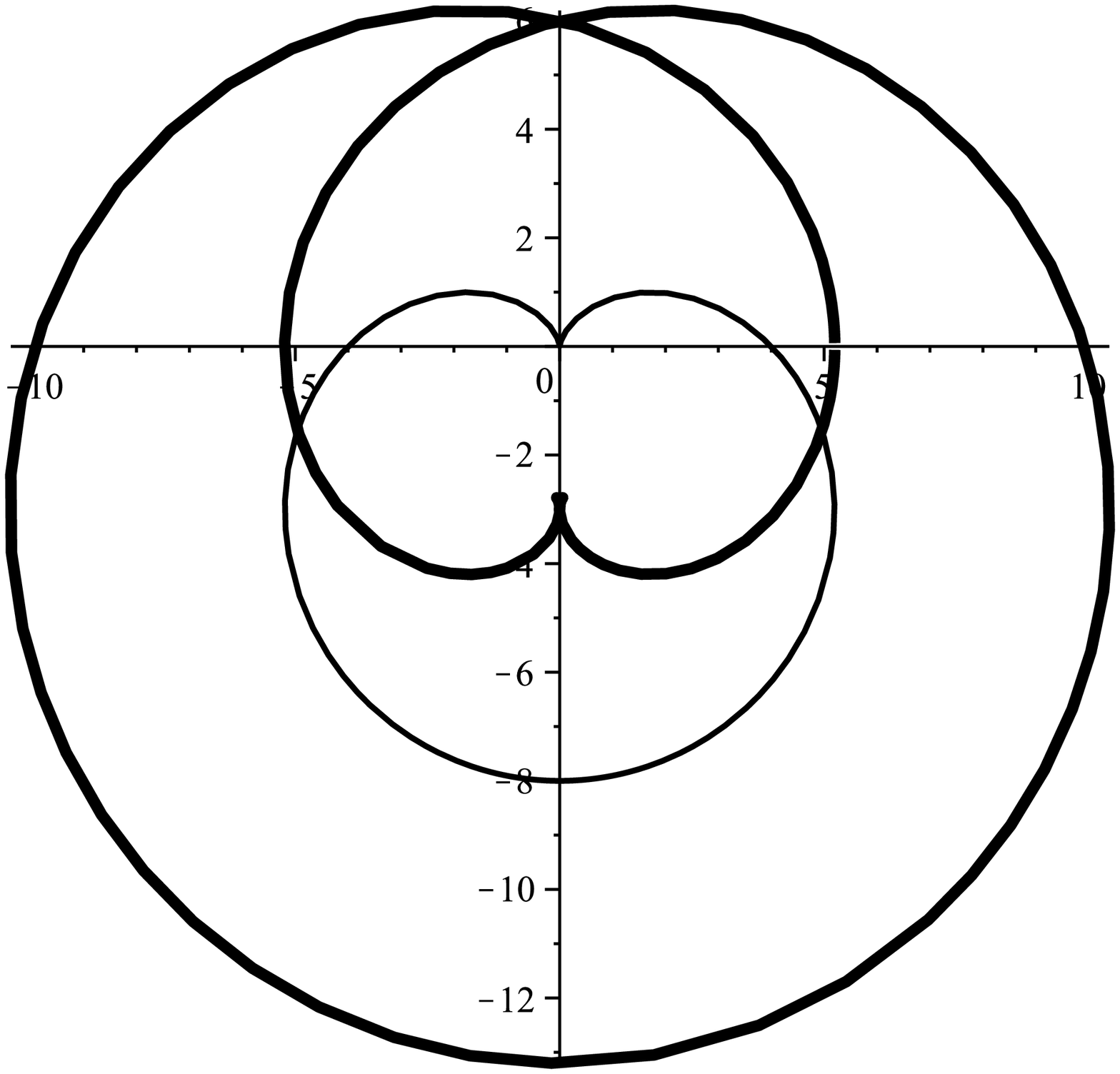,width=5cm,height=5cm} &
\psfig{figure=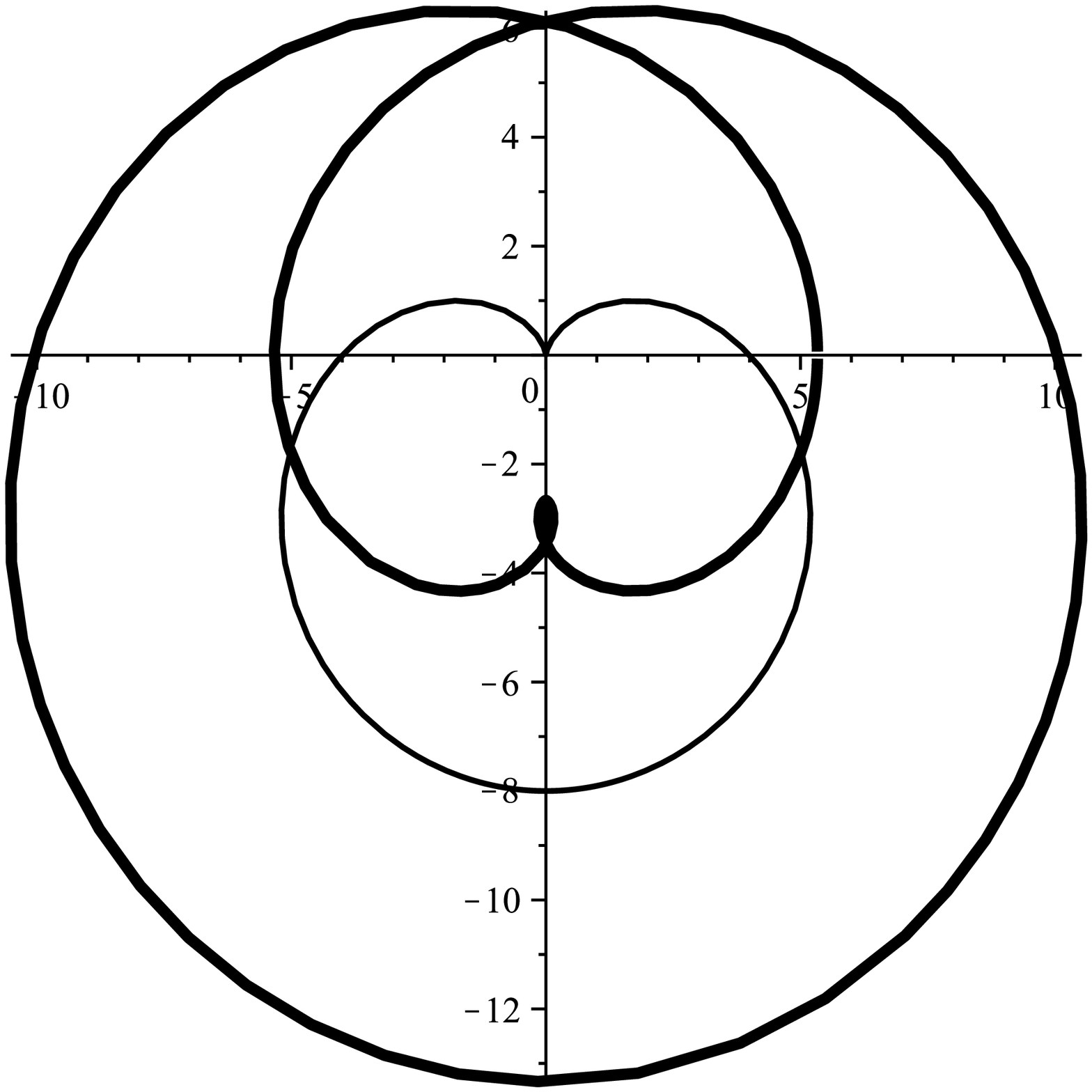,width=5cm,height=5cm}\\
& \psfig{figure=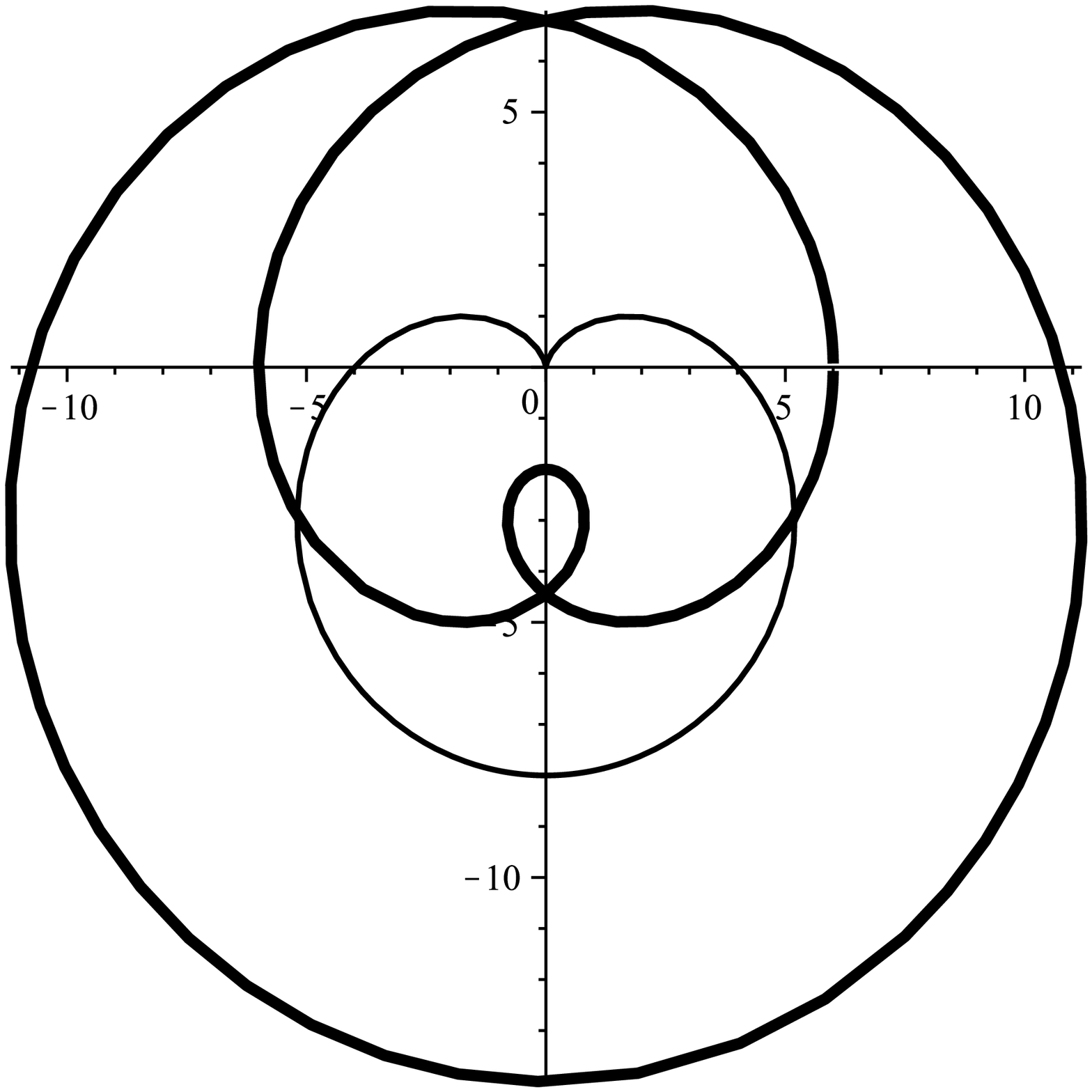,width=5cm,height=5cm} &\\
\end{array}$}
\end{center}
\caption{Offsets to the cardioid}
\end{figure}

\begin{figure}[ht]
\begin{center}
\centerline{$\begin{array}{ccc}
\psfig{figure=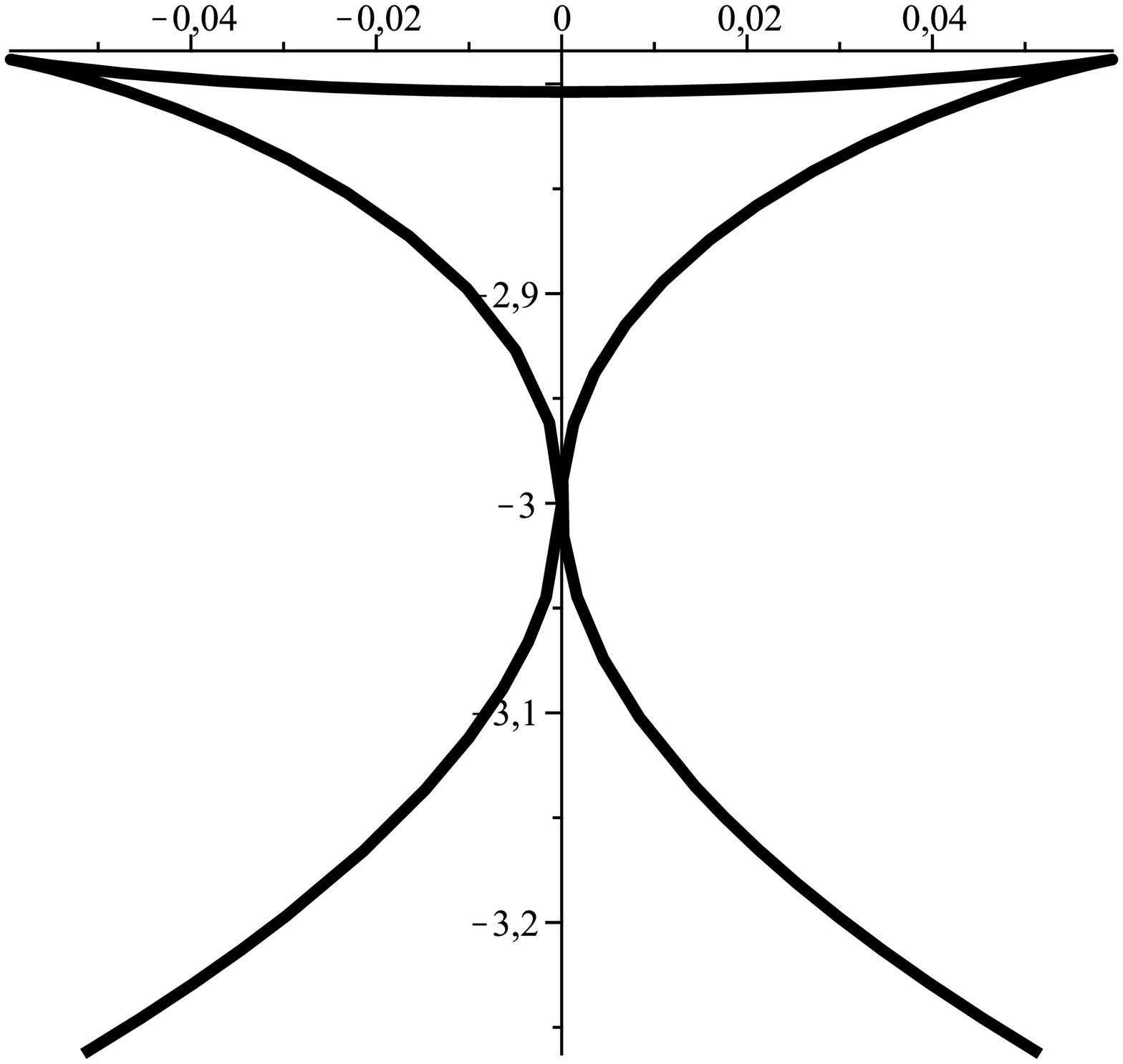,width=5cm,height=5cm} &
\psfig{figure=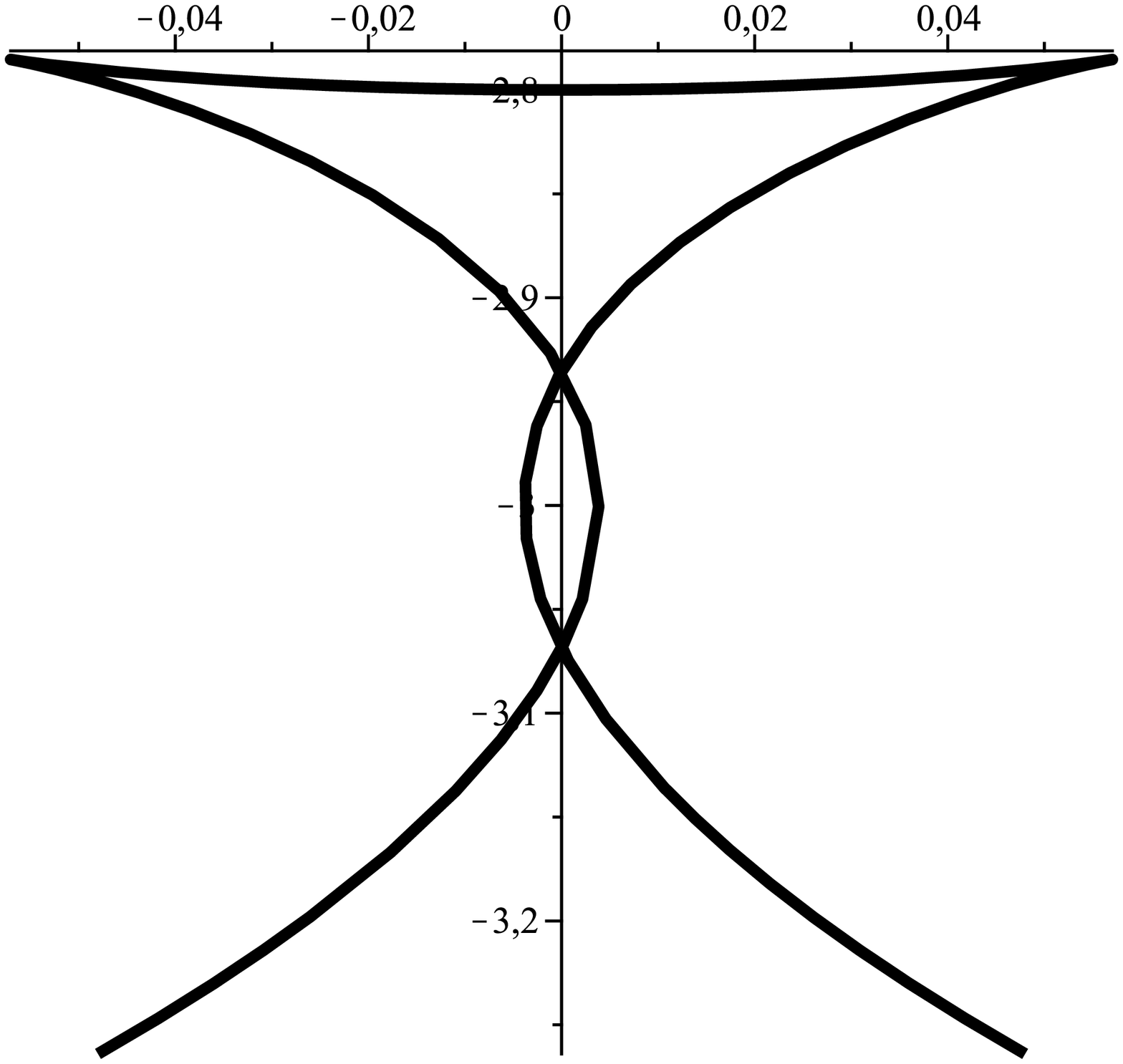,width=5cm,height=5cm} &
\psfig{figure=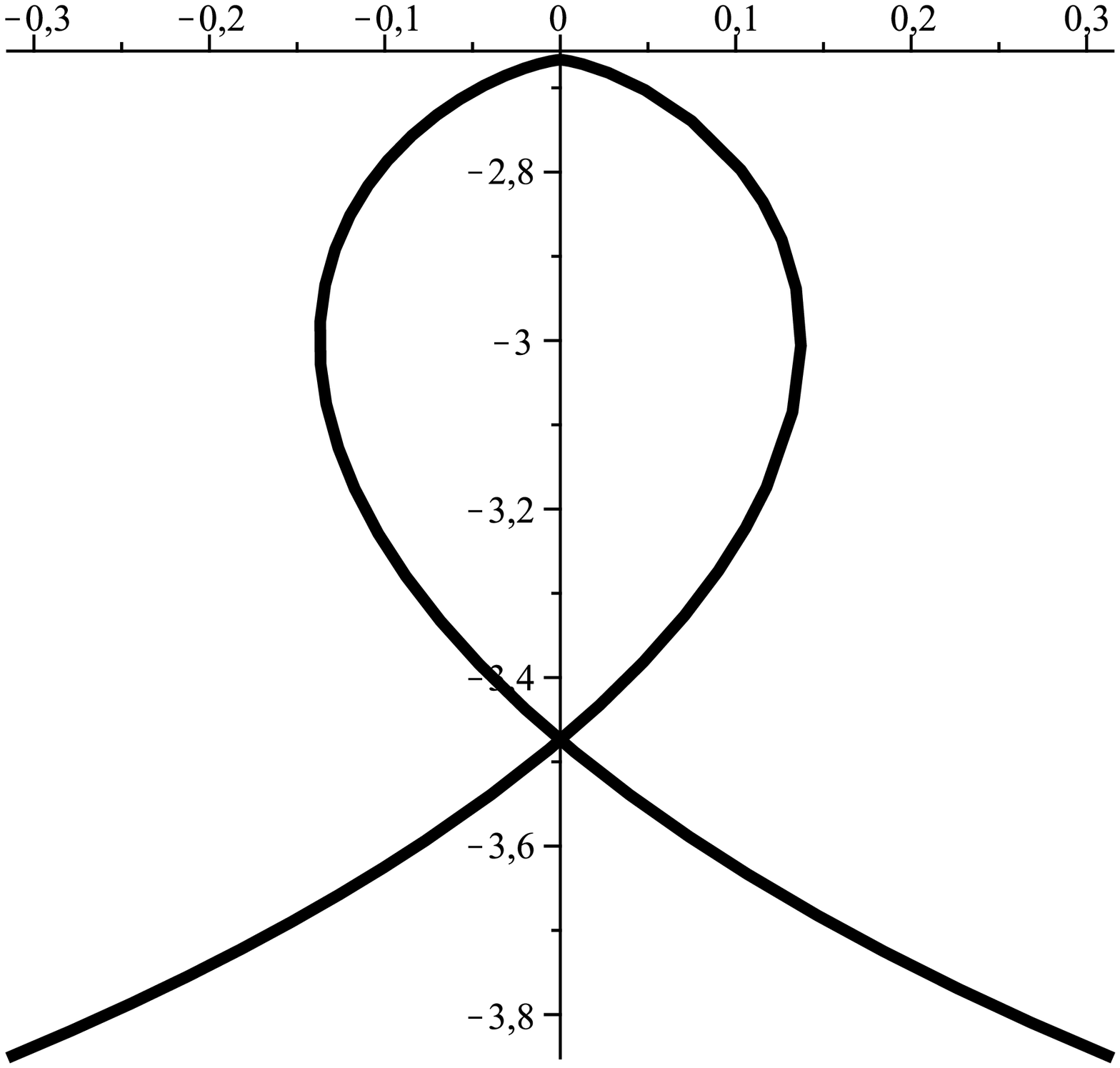,width=5cm,height=5cm}\\
\end{array}$}
\end{center}
\caption{Details of Some Offsets to the cardioid}
\end{figure}

{\bf Example 2:} {\it Consider the family
$\phi_d(t)=(u(t,d),v(t,d))$, where:
\[
\begin{array}{rcl}
u(t,d) &=& -25+11 t^2-29 t+d (57-95 t^2-22 t)\\
v(t,d) &= & 49+18 t^2+51 t+d (70+34 t^2-64 t)
\end{array}
\]
The implicit equation of the family is
$1136239-393995d-19629x-53165y+202885dx+130530992 d^3-1200232
d^2x+374269 d y-2090 d y^2+121 y^2+59360320 d^4+324 x^2-396
yx-33513124 d^2+1156 d^2 x^2+1224 d x^2-688992 d^2 y-1781936
yd^3+9025 d^2 y^2+2672 d x y+6460 y d^2 x-146992 x d^3=0$.

 Now by applying the algorithm in \cite{JGRS} we get that
 $\left\{\displaystyle{\frac{11}{25}}\right \}$ is a critical set. However,
 {\tt Critical} yields
 $\left \{\displaystyle{\frac{11}{25},\frac{-9}{17}}\right\}$ (which
 in this case coincides with ${\mathcal Spec}_1$, see Subsection \ref{subsec-summary});
 so, the second
 element of this last set is clearly superfluous. In fact, when plotting
 curves corresponding to $d\in (-\infty,11/25)$, $d=11/25$
 and $d\in (11/25,\infty)$, one gets a parabola in all the cases;
 hence, the topology type does not change for $d\in {\Bbb R}$,
 i.e. even the value $11/25$ provided by the implicit algorithm, is
 superfluous. The total amount of time required for the whole computation is 0.094 seconds.}

{\bf Comparison Table.} The following table shows a comparison
between the algorithm derived from the results in \cite{JGRS}, and
our algorithm, both of them implemented in Maple. For each family,
in this table we include whether the parametrization is rational
or polynomial (Type: R, rational, or P, polynomial), and we
provide the following data: the degree of the implicit equation of
the associated surface ($\deg(f)$), the total degree of the
parametrization ($\deg_t(\phi)$) i.e. the greatest power of the
parameter $t$ arising in the numerators and denominators of the
coordinates, the highest power ($\deg_{\lambda}(\phi)$) of
$\lambda$ arising in the numerators and denominators of the
coordinates, the timing for the algorithm in \cite{JGRS} (Imp., in
seconds), the timing for our algorithm (Crit., in seconds), the
size of the critical set determined by the parametric algorithm
(Size), and the difference (Dif.) between the sizes of the
critical sets provided by both algorithms (i.e. the size of the
critical set provided by {\tt Critical} minus the size of the
critical set provided by the algorithm in \cite{JGRS}). The symbol
* in the column of Imp. means that the algorithm has been unable
to provide an answer, or that the computation time exceeded
reasonable bounds. In Appendix I one may find the expressions of
all the parametrizations used here. It is worth mentioning that
except for the families numbers 3 and 7, the rest of the examples
have been randomly generated. Also, the timings given include the
cost of checking the hypotheses.

\begin{tabular}{|c|c|c|c|c|c|c|c|c|} \hline
$\mbox{Family}$ & $\mbox{Type}$ & $\deg(f)$ & $\deg_t(\phi)$ & $\deg_{\lambda}(\phi)$ & Imp. & Crit. & Size & Dif. \\
\hline \hline
1 & P. & 8 & 4 & 1 & 35.347 & 1.045 & 14 & 1 \\
2 & P. & 10 & 5 & 1 & 160.664 & 110.076 & 36 & 1 \\
3 & R. & 22 & 7 & 4 & * & 252.390 & 70 & * \\
4 & R. & 9 & 6 & 1 & * & 35.990 & 33 & * \\
5 & P. & 10 & 5 & 1 & 158.140 & 108.389 & 36 & 1 \\
6 & P. & 28 & 5 & 4 & 39.983 & 14.945 & 21 & 0 \\
7 & R. & 16 & 5 & 2 & * & 31.543 & 33 & * \\
8 & R. & 9 & 3 & 2 & 9.157 & 0.344 & 12& 3 \\
9 & R. & 24 & 6 & 3 & * & 74.770 & 78 & * \\
10 & R. & 6 & 2 & 2 & 0.109 & 0.093 & 7 & 2 \\
11 & R. & 12 & 3 & 3 & 19.580 & 0.389 & 25 & 0 \\
12 & R. & 18 & 3 & 3 & 66.253 & 1.325 & 30 & 9 \\
\hline
\end{tabular}

\subsection{Improvements in the computation.}\label{sec-additional}

The (1)-values, (3)-values and (2)-values not corresponding to
self-intersections of ${\mathcal M}$ can be more efficiently
computed by taking advantage of certain geometric properties of
${\mathcal C}$. In fact, one can determine these values by solving
polynomial systems in two variables; so, we can avoid one
resultant, and identify quite fast some values as potentially
critical. This is based on two classical results. The first one
follows essentially from Proposition 3 of \cite{GLMT}.

\begin{proposition} \label{sing-plane-space}
Let $Q=(x_q,z_q)$ be a singularity of ${\mathcal M}$, which is not a self-intersection of ${\mathcal M}$, and
such that $z_q$ is not a root of $\lcoeff_y(F)$. Then, one of these two possibilities occur: (i) $Q$ is the
projection of a singularity of ${\mathcal C}$; (ii) there exists a point of ${\mathcal C}$, projecting onto $Q$,
so that the tangent to ${\mathcal C}$ at this point is normal to the $xz$-plane.
\end{proposition}


The second result relates the non-singular points of ${\mathcal
M}$ with tangent parallel to the $z$-axis, to certain notable
points of ${\mathcal C}$. It can be easily proven by reasoning
with places.

\begin{proposition} \label{tangent}
Let $Q\in {\mathcal M}$ be a non-singular point of ${\mathcal M}$
with tangent parallel to the $x$-axis. Then, there exists some
point $Q'\in {\mathcal C}$, projecting onto $Q$, so that the
tangent to ${\mathcal C}$ at $Q'$ is parallel to the $xz$-plane.
\end{proposition}

So, let us consider first (1)-values and (2)-values. Those of
these values not corresponding to: (i) self-intersections of
${\mathcal M}$, (ii) real roots of $\lcoeff_y(F)$, (iii) points of
$\pi_{crit}({\mathcal C})$, are real $z_0$-values fufilling that
there exists $(x_0,z_0)\in {\mathcal M}_1\cup  {\mathcal M}_2$.
Now ${\mathcal M}_1$ can be seen as the union of the following two
curves: (1) the projection onto the $xz$-plane of the space curve
$\tilde{\mathcal C}_1$ defined by $f(x,t,z)=0,\mbox{ }h(t,z)=0$ in
the Euclidean space with coordinates $\{x,t,z\}$, which we denote
as $\pi_{xz}(\tilde{\mathcal C}_1)$; (2) the curve defined in the
$xz$-plane by $\gcd(\lcoeff_t(h),\lcoeff_t(f))$. The equation of
$\pi_{xz}(\tilde{\mathcal C}_1)$ is clearly $h(t,z)=0$. Thus, by
Proposition \ref{sing-plane-space} and Proposition \ref{tangent},
and using elementary properties of the resultant, one gets that
the considered values belonging to ${\mathcal M}_1$ also satisfy
$h_t(t,z)=0$, and hence they are contained in the set of real
roots of $\Res_t(h,h_t)$; one may observe that this set contains
also the real roots of $\gcd(\lcoeff_t(h),\lcoeff_t(f))$. Arguing
in a similar way for ${\mathcal M}_2$ we would reach the condition
$\Res_t(j,j_t)=0$. Moreover, the (3)-values can be related with
the asymptotes of the curves (in the $tz$-plane) $h(t,z)=0$ and
$j(t,z)=0$. So, the following theorem holds.

\begin{theorem} \label{simp-result}
The (1)-values, (2)-values not corresponding to self-intersections
of ${\mathcal M}$, and (3)-values not corresponding to asymptotes
of $\pi_{xz}({\mathcal C}_{crit})$, are among the finitely many
real roots of $\Res_t(h,h_t)$, $\Res_t(j,j_t)$.
\end{theorem}

\section{Conclusions} \label{sec-conclusions}

In this paper we have presented an algorithm for computing a
critical set of a family of rational curves depending on a
parameter. From the critical set, the topology types in the family
can be derived. The algorithm is based on a geometric
interpretation of known results for the implicit case, and on
advantages of parametric representation, and requires certain
properties on the family to be analyzed. These properties can be
algorithmically checked. In our experimentation, we have found
that the timings of the parametric algorithm are usually quite
better than those of the implicit algorithm; in fact, the
parametric algorithm is able to manage inputs that the implicit
algorithm cannot deal with. On the other hand, the drawback of the
provided algorithm is that it may determine critical sets bigger
than those determined by the implicit algorithm, therefore
containing superfluous values with respect to the implicit
critical set. So, as a potential future line of research, one
could address the problem of reducing the size of the output,
trying to approach optimality. Furthermore, the method applies
with exact coefficients. So, it would also be nice to consider the
(challenging) possibility of applying it in the case of
approximate coefficients.

{\bf Acknowledgements.} The author wants to thank the referees of
the paper for their useful observations and suggestions, that
helped to improve the paper.

\section{Appendix I: Parametrizations of the families used in the comparison table.} \label{appendix}

{\bf Family 1:}
\[
\begin{array}{rcl}
u &:=& -78t^4+62t^3+11t^2+88t+1+\lambda(30t^4+81t^3-5t^2-28t+4)\\
v &:=& -11t^4+10t^3+57t^2-82t-48+\lambda(-11t^4+38t^3-7t^2+58t-94)
\end{array}
\]
The implicit equation has degree 4 (as a polynomial in $x,y$) and
95 terms; the total degree can be found (for this family and for
the rest of the families in this appendix) in the comparison table
provided in Subsection \ref{sec-param-initial}.

{\bf Family 2:}
\[
\begin{array}{rcl}
u &:=& 50-85t^5-55t^4-37t^3-35t^2+97t+\lambda(-59+79t^5+56t^4+49t^3+63t^2+57t)\\
v &:=& -62+45t^5-8t^4-93t^3+92t^2+43t+\lambda(-61+77t^5+66t^4+54t^3-5t^2+99t)
\end{array}
\]
The implicit equation has degree 5 (as a polynomial in $x,y$) and
161 terms.

{\bf Family 3:}
\[
\begin{array}{rcl}
u &:=& \displaystyle{\frac{2t^7+t^5\lambda+\lambda^2t^4-3t^3\lambda^2+3\lambda^3t^2-t^3+3t\lambda-2\lambda^2-2t^4\lambda+t^6\lambda-t^4}{(t^4-2t\lambda+\lambda^2)(t^3-t\lambda+\lambda^2)}}
\\
v &:=& \displaystyle{\frac{t^3+\lambda t^2-1}{t^3-t \lambda + \lambda^2}}
\end{array}
\]
The implicit equation has degree 7 (as a polynomial in $x,y$) and
343 terms.

{\bf Family 4:}
\[
\begin{array}{rcl}
u &:=& \displaystyle{\frac{-35-85t^3-55t^2-37t}{56+97t^3+50t^2+79t}}+\lambda\displaystyle{\frac{66+43t^3-62t^2+77t}{-61+54t^3-5t^2+99t}}
\\
v &:=& \displaystyle{\frac{31-50t^3-12t^2-18t}{56+97t^3+50t^2+79t}}+\lambda\displaystyle{\frac{-59+49t^3+63t^2+57t}{-61+54t^3-5t^2+99t}}
\end{array}
\]
The implicit equation has degree 6 (as a polynomial in $x,y$) and
84 terms.

{\bf Family 5:}
\[
\begin{array}{rcl}
u &:=& 50-85t^5-55t^4-37t^3-35t^2+97t+\lambda(-59+79t^5+56t^4+49t^3+63t^2+57t)\\
v &:=& -62+45t^5-8t^4-93t^3+92t^2+43t+\lambda(-61+77t^5+66t^4+54t^3-5t^2+99t)
\end{array}
\]
The implicit equation has degree 5 (as a polynomial in $x,y$) and
161 terms.

{\bf Family 6:}
\[
\begin{array}{rcl}
u & :=& -47t-91\lambda^2-47t^3-61\lambda^4+41t^5-58t^2\lambda^3\\
v &:=& 23t^2-84t^3\lambda+19t^2\lambda^2-50t\lambda^3+88t^5\lambda-53t^2\lambda^4
\end{array}
\]
The implicit equation has degree 5 (as a polynomial in $x,y$) and
191 terms.

{\bf Family 7:}
\[
\begin{array}{rcl}
u &:=& \displaystyle{\frac{t^5-t^2\lambda^2-t-2\lambda+1}{t^3-t^2+t\lambda-\lambda^2}}\\
v &:=& \displaystyle{\frac{t^5+t^2\lambda-t-2\lambda^2+1}{t^3-t^2+t\lambda-\lambda^2}}
\end{array}
\]
The implicit equation has degree 5 (as a polynomial in $x,y$) and
219 terms.

{\bf Family 8:}
\[
\begin{array}{rcl}
u & := & \displaystyle{\frac{-7t+58t^2-94t\lambda-68t^3+14t^2\lambda-35\lambda^3}{-14-9t-51\lambda-73t^2-73t\lambda-91\lambda^2}}\\
v & := & \displaystyle{\frac{-50+50\lambda+67t^2-39t\lambda+8\lambda^2-49t\lambda^2}{-14-9t-51\lambda-73t^2-73t\lambda-91\lambda^2}}
\end{array}
\]
The implicit equation has degree 3 (as a polynomial in $x,y$) and
72 terms.

{\bf Family 9:}
\[
\begin{array}{rcl}
u & :=& \displaystyle{\frac{-5+99t-61\lambda-50\lambda^3-12t^6-18\lambda^6}{31-26t-62\lambda+t^2-47t\lambda-91\lambda^2}}\\
v & := & -1+94t^2+83\lambda^2-86t\lambda^2+23\lambda^3-84t^3\lambda
\end{array}
\]
The implicit equation has degree 6 (as a polynomial in $x,y$) and
204 terms.

{\bf Family 10:}
\[
\begin{array}{rcl}
u&:=&\displaystyle{\frac{-85-55t-37\lambda-35t^2+97t\lambda+50\lambda^2}{79+56t+49\lambda+63t^2+57t\lambda-59\lambda^2}}\\
v&:=&\displaystyle{\frac{45-8t-93\lambda+92t^2+43t\lambda-62\lambda^2}{79+56t+49\lambda+63t^2+57t\lambda-59\lambda^2}}
\end{array}
\]
The implicit equation has degree 2 (as a polynomial in $x,y$) and
30 terms.

{\bf Family 11:}
\[
\begin{array}{rcl}
u&:=&\displaystyle{\frac{97\lambda+50t\lambda+79\lambda^2+56t^3+49t\lambda^2+63\lambda^3}{-93t+92\lambda+43t\lambda-62t^3+77t\lambda^2+66\lambda^3}}\\
v&:=&\displaystyle{\frac{-12-18t+31\lambda-26t\lambda-62\lambda^2+t^2\lambda}{-93t+92\lambda+43t\lambda-62t^3+77t\lambda^2+66\lambda^3}}
\end{array}
\]
The implicit equation has degree 3 (as a polynomial in $x,y$) and
97 terms.

{\bf Family 12:}
\[
\begin{array}{rcl}
u&:=&\displaystyle{\frac{57t-59t\lambda+45\lambda^2-8t^3-93t\lambda^2+92t^2\lambda^2}{-18t+31t^2-26t\lambda-62t^3+t^2\lambda^2-47\lambda^4}}\\
v&:=&\displaystyle{\frac{-1+94t^2+83\lambda^2-86t\lambda^2+23\lambda^3-84t^3\lambda}{-18t+31t^2-26t\lambda-62t^3+t^2\lambda^2-47\lambda^4}}
\end{array}
\]
The implicit equation has degree 3 (as a polynomial in $x,y$) and
146 terms.


\begin{thebibliography}{56}

\bibitem{JGRS} Alcazar J.G., Schicho J., Sendra R. (2007) {\it A
Delineability-based Method for Computing Critical Sets of Algebraic Surfaces}, Journal of Symbolic Computation vol. 42, pp.
678-691









\bibitem{AC07} Andradas C., Recio T. (2007) {\it Plotting missing points and branches of real parametric curves},
Applicable Algebra in Engineering and Computing 18 (1-2), pp. 107-126.

\bibitem{ASS96} Arrondo E., Sendra J., Sendra J.R. (1997).
{\it  Parametric Generalized Offsets to Hypersurfaces}.   Journal
of Symbolic Computation vol. 23, pp. 267--285.

\bibitem{Basu} Basu S., Pollack R., Roy M.F. (2003)  {\it Algorithms in
Real Algebraic Geometry} , Springer Verlag.


\bibitem{Eigen} Eigenwilling A., Kerber M., Wolpert N. (2007) {\it Fast and Exact Geometric Analysis of Real Algebraic Plane Curves}, in C.W. Brown, editor, Proc. Int. Symp. Symbolic and Algebraic Computation, pp. 151-158, Waterloo, Canada. ACM.


\bibitem{Far1} Farouki R.T., Neff C.A. (1990). {\it Analytic
Properties of Plane Offset Curves}, Computer Aided Geometric
Design 7, pp. 83--99.


\bibitem{GLMT} Gatellier G., Labrouzy A., Mourrain B.,
Tecourt J.P. (2004) {\it Computing the topology of
three-dimensional algebraic curves}. Computational Methods for
Algebraic Spline Surfaces, pages 27-44. Springer-Verlag.


\bibitem{Lalo} Gonzalez-Vega L., Necula I. (2002).
{\it Efficient topology determination of implicitly defined
algebraic plane curves}, Computer Aided Geometric Design, vol. 19
pp. 719-743.





\bibitem{Hong} Hong H. (1996). {\it An effective
method for analyzing the topology of plane real algebraic curves},
Math. Comput. Simulation 42 pp. 571-582



\bibitem{LaloCompl} Gonzalez-Vega L.,   El
Kahoui M. (1996). {\it  An improved upper complexity bound for the
topology computation of a real algebraic plane curve}, J.
Complexity 12 pp 527-544.


\bibitem{L95a} L\"u W.  (1995),   {\it Offset-Rational Parametric
Plane Curves},   Computer Aided Geometric Design {\bf 12},
601-617.

\bibitem{Mourrain-2} Mourrain B., Tecourt J. (2005), {\it Isotopic
Meshing of a Real Algebraic Surface}, Rapport de recherche nº 5508,
Unite de Recherche INRIA Sophia Antipolis.


\bibitem{PD07} P\'erez-D\'{\i}az S. (2007), {\it Computation of the singularities of parametric plane curves}, Journal of Symbolic Computation 42, pp. 835-857.

\bibitem{Po95} Pottmann H.  (1995), {\it Rational Curves and
Surfaces with Rational Offsets}. Computer Aided Geometric Design
{\bf 12}, 175-192.

\bibitem{PP98b} Pottmann H., Peternell M. (1998), {\it
A Laguerre Geometric Approach to Rational Offsets}.  Computer
Aided Geometric Design {\bf 15/3},  223-249.




\bibitem{Juani-thesis} Sendra J. (1999) {\it Algoritmos efectivos para la
manipulacion de offsets de hipersuperficies}, PhD Thesis,
Universidad Politecnica de Madrid.

\bibitem{S02} Sendra J.\,R. (2002).
{\it Normal Parametrizations of Algebraic Plane Curves}.
 Journal of Symbolic Computation vol. 33, pp. 863--885.




\bibitem{SWPD} Sendra J.R., Winkler F., Perez-Diaz P. (2008). {\it Rational Algebraic Curves}, Springer-Verlag.

\bibitem{Shafa} Shafarevich, I.R. (1994). {\it Basic Algebraic Geometry}, Springer-Verlag.

\bibitem{walker} Walker R.\,J. (1950). {\it  Algebraic Curves.}
Princeton University Press, Princeton.

\bibitem{Wi96} Winkler F. (1996), {\it Polynomial Algorithms in Computer Algebra}.
Springer Verlag, ACM Press.

\end{thebibliography}
\end{document}